\documentclass[usenatbib]{mnras}

\usepackage{graphicx}	% Including figure files
\usepackage{amsmath}	% Advanced maths commands
\usepackage{amssymb}	% Extra maths symbols

\usepackage{txfonts}
\usepackage{tabularx}

\usepackage[T1]{fontenc} % for special font in bibliography e.g., {S{\k{a}}dowski}
\usepackage[usenames,dvipsnames,svgnames,table]{xcolor}

\newcommand\TopTabspace{\rule{0pt}{2.6ex}}       % Vertical Space of table (Top) 
\newcommand\BotTabspace{\rule[-1.4ex]{0pt}{0pt}} % Vertical Space of table (Bottom)

\newcommand{\hammer}{{\tt{H-AMR}}~}
\newcommand{\harm}{{\tt{HARM2D}}~}

\newcommand{\bhoss}{{\tt{BHOSS}}~}
\newcommand{\grmonty}{{\tt{GRMONTY}}~}

\newcommand{\eht}{{\tt{EHT}}~}
\newcommand{\gravity}{{\tt{GRAVITY}}~}

\newcommand{\sgra}{Sgr~A$^{*}$}

\newcommand{\eq}[1]{Eq.~(\ref{#1})}
\newcommand{\fig}[1]{Figure~\ref{#1}}
\newcommand{\MsunYr}{M_{\odot}\,{\rm yr}^{-1}}

\title[Spectra of \sgra{} from 3D radiative GRMHD]{Spectral and Imaging properties of \sgra{} from High-Resolution 3D GRMHD Simulations with Radiative Cooling}

\author[Yoon et al.]
{D. Yoon$^{1}$\thanks{E-mail: d.yoon@uva.nl}, 
K. Chatterjee$^{1}$, 
S.B. Markoff$^{1,2}$, 
D. van Eijnatten$^{1}$, 
Z. Younsi$^{3,4}$, 
\newauthor 
M. Liska$^{5,1}$ 
\& A. Tchekhovskoy$^{6}$   
\\
$^{1}$Anton Pannekoek Institute for Astronomy, University of Amsterdam, Science Park 904, 1098 XH Amsterdam, The Netherlands\\
$^{2}$Gravitation and Astroparticle Physics Amsterdam (GRAPPA) Institute, University of Amsterdam, Science Park 904, 1098 XH Amsterdam, The Netherlands\\
$^{3}$Mullard Space Science Laboratory, University College London, Holmbury St.~Mary, Dorking, Surrey, RH5 6NT, United Kingdom\\
$^{4}$Institut f{\"u}r Theoretische Physik, Goethe-Universit{\"a}t Frankfurt, Max-von-Laue-Stra{\ss}e 1, D-60438 Frankfurt am Main, Germany\\
$^{5}$Institute for Theory and Computation, Harvard University, 60 Garden Street, Cambridge, MA 02138, USA; John Harvard Distinguished Science and ITC\\ Fellow\\
$^{6}$Center for Interdisciplinary Exploration \& Research in Astrophysics (CIERA), Physics \& Astronomy, Northwestern University, Evanston, IL 60202, USA\\
}

\date{Accepted XXX. Received YYY; in original form ZZZ}

\pubyear{2020}

\begin{document}
\label{firstpage}
\pagerange{\pageref{firstpage}--\pageref{lastpage}} 
\maketitle

\begin{abstract}
    The candidate supermassive black hole in the Galactic Centre, Sagittarius A* (\sgra{}), is known to be fed by a radiatively inefficient accretion flow (RIAF), inferred by its low accretion rate. Consequently, radiative cooling has in general been overlooked in the study of \sgra{}. However, the radiative properties of the plasma in RIAFs are poorly understood. In this work, using full 3D general-relativistic magneto-hydrodynamical simulations, we study the impact of radiative cooling on the dynamical evolution of the accreting plasma, presenting spectral energy distributions and synthetic sub-millimeter images generated from the accretion flow around \sgra{}. These simulations solve the approximated equations for radiative cooling processes self-consistently, including synchrotron, bremsstrahlung, and inverse Compton processes. 
    %We extend previous parameter study in two-dimensional simulations to full three-dimension, which is essential to sustain the steady accretion flows via magneto-rotational instability-driven turbulence.
    %To enable practical calculations of cooling in GRMHD simulations, we adopt an electron temperature model, at which the temperature ratio of ions and electrons, $T_i/T_e$, is constant or depends on the local magnetic dominance. This model assumes that energy transfer from ions to electrons is fast enough to keep the temperature ratio constant. 
    We find that radiative cooling plays an increasingly important role in the dynamics of the accretion flow as the accretion rate increases: the mid-plane density grows and the infalling gas is less turbulent as cooling becomes stronger. The changes in the dynamical evolution become important when 
    %We also find that while cooling may be less effective for the lowest accretion rate limit of \sgra{}, radiative cooling can drive significant changes in the dynamical evolution when 
    the accretion rate is larger than $10^{-8}\,M_{\odot}~{\rm yr}^{-1}$ ($\gtrsim 10^{-7} \dot{M}_{\rm Edd}$, where $\dot{M}_{\rm Edd}$ is the Eddington accretion rate). The resulting spectra in the cooled models also differ from those in the non-cooled models: the overall flux, including the peak values at the sub-mm and the far-UV, is slightly lower as a consequence of a decrease in the electron temperature. Our results suggest that radiative cooling should be carefully taken into account in modelling \sgra{} and other low-luminosity active galactic nuclei that have a mass accretion rate of $\dot{M} > 10^{-7}\,\dot{M}_{\rm Edd}$.
\end{abstract}
    %and other low-luminosity active galactic nuclei that have a mass accretion rate of $\dot{M} > 10^{-7}\,\dot{M}_{\rm Edd}$.
	%Since the mass accretion rate is far below the maximum Eddington limit in \sgra{},
	%Depending on how low the accretion rate of \sgra{} is, these effects may not be significant for the lowest accretion limit, however, radiative cooling leads to significant changes in the dynamical evolution already when the accretion rate is larger than $10^{-8}\,\MsunYr$. 

%%%%%%%%%%%%%%%%%%%%%%%%%%%%%%%%%%%%%%%%%%%%%%%%%%

%%%%%%%%%%%%%%%%% BODY OF PAPER %%%%%%%%%%%%%%%%%%

\section{Introduction}
\label{sec:intro}

It is widely believed that most galaxies harbour supermassive black holes (SMBHs) in their galactic centres, with masses ranging from millions to billions of solar masses. Over the past few decades, the black hole (BH) candidate in the centre of the Milky Way, Sagittarius A* (hereafter \sgra{}), has proven an exceptional laboratory for studies of accretion and outflow physics of BHs due to its proximity to Earth. A significant effort has been invested in determining the BH mass and distance for \sgra{} \citep[e.g.,][]{Reid:93, Reid:19, Schodel:02,Bower:04,Ghez:08,Gillessen:09,Gillessen:17,Boehle:16, Gravity:18}. We adopt the current best-fit BH mass of $4.1\pm0.03 \times 10^6\, M_{\odot}$, which was measured by the orbital motion of stars and gas clouds \citep{Gillessen:17, Gravity:18}, and the distance of $8.15\pm0.15$ kpc, which was obtained from trigonometric parallaxes and proper motions of massive stars around \sgra{} \citep{Reid:19}. Given the mass and the distance, the angular size of the Schwarzschild radius, $r_{\rm S} = 2GM_{\rm BH}/c^2$, is $\approx 10\,\mu\rm as$, which subtends a larger area in the sky than any other known BH, including all stellar-mass BHs.

Recently, mounting attention has been paid to the study of \sgra{} with the advent of the pioneering instruments \gravity{} \citep{Gravity:17, Gravity:18} and the {\tt Event Horizon Telescope} (\eht{}; \citealt{Doeleman:09,EHT:19b}), capable of probing 10 -- 30 $\mu\rm as$ scales.  These instruments allow us to profoundly improve our understanding of the physical processes associated with the accretion and relativistic jet formation in the immediate vicinity of SMBHs and thus demand equal measures of theoretical support and precise modelling of the spectrum generated by the radiation from the accretion flow around \sgra{}.

%This angular resolution is comparable to the size of the Schwarzschild radius of \sgra{}.

%\cmtdy{add multiwavelength spectrum}

The mass accretion rate around \sgra{} (in units of $\MsunYr$) is estimated to be in the range of $\sim10^{-9} < \dot{M} < 10^{-7}$, as constrained by the measured Faraday rotation measure at mm/sub-mm wavelengths \citep{Aitken:00,Bower:03,Marrone:07}. Such a low accretion rate favours hot accretion flow models for the accretion disk instead of the radiatively efficient, thin disk models \citep{Shakura:73}. Many theoretical scenarios have been invoked and excluded to account for the nature of accretion and outflows in the hot accretion flow: the {\it standard} advection dominated accretion flow (ADAF; \citealt{Narayan:94,Narayan:95a,Narayan:98}) and Bondi-Hoyle models are ruled out, since these models are expected to yield an accretion rate of $\sim 10^{-5}\,\MsunYr$, which is two orders of magnitude higher than the measured upper limit \citep{Bower:03}.  \citet{Yuan:03} reexamined the radiatively inefficient accretion flow (RIAF) model for the spectrum of \sgra{} and argued that the presence of outflows within the Bondi radius plays a vital role in reducing the mass accretion rate. Alternatively, the convection dominated accretion flow (CDAF) and jet-dominated models are capable of reproducing the spectrum that is consistent with the observed accretion rate \citep{Quataert:00, Falcke:00, Markoff:07}. The existence of such outflows is supported by the weak hydrogen-like Fe K$\alpha$ line around \sgra{} via the X-ray Visionary Program \citep{Wang:13}. In this study, the flat density profile in the spectrum confirmed that $\gtrsim 99\%$ of the matter initially captured by the SMBH is lost before it reaches the innermost region around \sgra{}, which is consistent with the CDAF model or adiabatic inflow-outflow solution (ADIOS; see \citealt{Blandford:99,Yuan:14} for the detailed model descriptions) model. 

%\citep[see][references therein]{Yuan:14}.

Although  semi-analytic models provide an important framework for understanding the nature of the accretion flow around \sgra{}, numerical simulations are required to capture the time-dependent, turbulent evolution of the accretion flow. In particular, a self-consistent magneto-hydrodynamical (MHD) description enables us to demonstrate accretion processes induced by the magneto-rotational instability (MRI; \citealt{Balbus:91}) without imposing an arbitrary anomalous viscosity. In earlier numerical studies, three-dimensional pseudo-Newtonian MHD simulations were carried out to model the synchrotron radiation from accretion flows\citep[e.g.,][]{Goldston:05,Ohsuga:05,Chan:09,Huang:09}.  However, the non-relativistic treatment in the simulations has disadvantages for modelling the synchrotron radiation, since it is mainly emitted in the immediate vicinity of the central BH, where relativistic effects cannot be ignored: shocks develop differently for the relativistic plasma when subject to intense magnetic and gravitational fields \citep{DelZanna:03}, and the curvature of space-time becomes significant. Several other works made use of general-relativistic magneto-hydrodynamical (GRMHD) simulations for studying the dynamics and spectral properties of \sgra{} in two dimensions \citep[2.5D; e.g,][]{Noble:07,Moscibrodzka:09,Hilburn:10,Dibi:12,Drappeau:13,Moscibrodzka:13}, or in three dimensions (3D) \citep[e.g.,][]{Dexter:09,Dexter:10,Dolence:12,Shcherbakov:12,Dexter:13,Moscibrodzka:14,Davelaar:18,Chael:18}.  In general, 2.5D simulations are a reasonable and computationally cheaper option to conduct a parameter study for reproducing the spectrum of \sgra{}, but it is known that simulations with axisymmetric coordinates cannot sustain MRI-driven turbulence, which decays over the local orbital time as a consequence of Cowling's anti-dynamo theorem (\citealt{Cowling:33}, see also \citealt{Hide:82} for the generalised description). Therefore, full 3D GRMHD simulations are necessary to perform detailed studies of the nature of the accretion flows around \sgra{} and the emitted spectrum. For instance, it was confirmed that thick accretion disks are able to generate and advect large-scale poloidal magnetic flux through dynamo action when resolved properly \citep{Liska:18b}. 

%Goldston, Quataert & Igumenshchev 2005; Ohsuga, Kato & Mineshige 2005; Dexter, Agol & Fragile
%2009; Mo´scibrodzka et al. 2009; Dexter et al. 2010; Hilburn et al. 2010; Shcherbakov, Penna &
%McKinney 2010; Dexter & Fragile 2012; Dolence et al. 2012; Shiokawa et al. 2012)

The bolometric luminosity of \sgra{} is extremely low, $L_{\rm bol} \sim 10^{36}\, {\rm erg\,s^{-1}} \approx 10^{-9}\,L_{\rm Edd}$, where $L_{\rm Edd}$ is the Eddington luminosity. Given such a low luminosity, it has been thought that the radiative cooling losses of \sgra{} are negligible, since the losses are likely not strong enough to have a significant impact on the dynamics of the accretion flow. Based on this argument, all previous works with full 3D GRMHD simulations have ignored the radiative cooling losses for simplicity. Although this assumption may be reasonable, \citet{Dibi:12} argued based on their 2.5D simulations that cooling losses play an increasingly important role for higher accretion rates and possibly alter the dynamics and resulting spectra of \sgra{}, even within the allowed range of accretion rates based on polarisation and X-ray studies. One potential impact is that the radiative cooling reduces the gas pressure and the disk vertical scale height, resulting in a decrease in turbulence and a more ordered magnetic field \citep{Fragile:09}. Moreover, many questions remain unanswered: how do radiative cooling losses affect the turbulence features of the disk, and thus the angular momentum transfer of the accreting plasma? How does radiative cooling together with GR effects result in the observed spectra from \sgra{}? Is radiative cooling {\it indeed} negligible for the mass accretion rate range of \sgra{}? Even though the effects of radiative cooling can be minor for the case of \sgra{}, the quantitative evaluation of cooling effects is highly demanded because it must play a greater role for SMBHs with higher mass accretion rates, such as M87.

In this paper, we perform the first full 3D GRMHD simulations which include cooling losses via bremsstrahlung, thermal synchrotron emission, and inverse Compton scattering. Due to the significant computational expense of the full 3D simulations, we cannot explore the full range of various parameters (e.g., spin, magnetic configuration, electron distribution function, misaligned disk). Instead, we use parameters compatible with earlier studies, assuming a rapidly rotating BH, weak poloidal initial magnetic field, and a fixed temperature ratio between ions and electrons of $T_{i}/T_{e}=3$ (we also carry out additional simulations with different electron temperature prescriptions for comparison). We then examine the effect of radiative cooling on the dynamics of the accretion flow and the resulting spectra and images, for different accretion rates within the allowed range.

%\cmtdy{tilted disk, non-thermal electron distribution ... for following studies}\cmtkc{maybe add these in discussion as future work}

This paper is structured as follows. In \S\,\ref{sec:tech}, we give a technical description of the numerical methods used, including the simulation setup, and the treatment of radiative cooling losses. In \S\,\ref{sec:results}, we describe the results of how cooling losses play a role in changing the dynamical evolution of accreting matter. In \S\,\ref{sec:discuss} we discuss the best-bet model for \sgra{}, the effects of cooling on the resulting spectra and sub-mm images, and the variability of multi-wavelength spectra. We also compare our 3D work to previous 2.5D work. We summarise our results in \S\,\ref{sec:conclusions}.

\section{Technical description of method}
\label{sec:tech}

All simulations are performed with the \hammer code \citep{Liska:19b,Porth:19}, which branched off \harm \citep{Gammie:03,Noble:06} in its early days.  It is accelerated by Graphical Processing Units (GPUs) and improved with a staggered grid for constrained transport of magnetic fields \citep{Gardiner:05} to preserve $\nabla \cdot B = 0$, more robust inversion \citep{Newman:14} adaptive mesh refinement (AMR, not utilised in this work), static mesh refinement (SMR), and a locally adaptive time step \citep[LAT; see][ Appendix A]{Chatterjee:19}. It adopts a piece-wise parabolic method (PPM; \citealt{Colella:84}) for reconstruction of cell-centred quantities at cell faces, which is third-order accurate, for the spatial reconstruction at cell faces from cell centres, and a second-order time-stepping.

The broadband spectrum is calculated from the GRMHD output, using the general-relativistic Monte Carlo scheme \grmonty \citep{Dolence:09}, which includes synchrotron emission and absorption, and inverse Compton scattering for a relativistic thermal Maxwell-J\"uttner distribution of electrons. Technically, \grmonty{} cannot produce synthetic images but only spectra. Thus, we ray-trace the GRMHD-produced spectra by integrating the general-relativistic radiative transfer (GRRT) equations using the \bhoss code \citep{Younsi:12,Younsi:16,Younsi:19_polarizedbhoss} to generate synthetic images at 230 GHz that can help us infer the expected images of \sgra{} from the upcoming \eht{} project. In \bhoss{}, the calculation of radiative processes includes synchrotron emission and absorption only, which is sufficient for imaging at the EHT frequency of 230 GHz. Since the sub-mm regime of the spectra are dominated by synchrotron emission, which both codes calculate, we verify consistency in our spectral calculations from both codes by comparing the resulting spectra in the radio to NIR bands (see Appendix~\ref{sec:GRRT}). 

%We also generate infrared synthetic images using the
%general-relativistic ray-tracing code \bhoss (Younsi et al. 2019, in prep.) which includes
%synchrotron emission/absorption. Note that the emission is calculated in post-processing, and thus
%the effects of radiation forces acting on the gas during the dynamical evolution are not taken into
%account.

\subsection{Numerical Grid and Floors}
\label{subsec:grid}

For convenience, we adopt Heaviside-Lorentz units, which absorb a factor of $\sqrt{4\pi}$ for the magnetic field 4-vector, $b^{\mu}$, so that the magnetic pressure is $P_{\rm B}\equiv b^2/2$. Furthermore, the typical natural units are used, $GM=c=1$, which sets the length unit to be the gravitational radius, $r_{\rm g} \equiv G\,M / c^2$, and  the time unit to be the light crossing time, $t_{\rm g} \equiv G\,M / c^3$, where $G,\,M,\,c$ are the gravitational constant, BH mass, and the speed of light, respectively.  We use a spherical-polar axisymmetric computational grid ($r,\,\theta,\,\phi)$ extending from 0.85 $r_{\rm H}$ to 250 $r_{\rm g}$, where the event horizon radius $r_{\rm H} \equiv r_{\rm g} \left( 1+\sqrt{1-a_\star^2} \right)$. Here we set the dimensionless BH spin parameter to $a_\star \equiv c\,J / G M^2$ in a Kerr-Schild foliation, where $J=r_{\rm H}\,c^2$ is the angular momentum at the event horizon. The grid is uniformly spaced with respect to a set of internal coordinates $\left(x^{1},\, x^{2},\, x^{3}\right)$, which can be converted to ($r,\,\theta,\,\phi)$, respectively \textbf{\footnote{The coordinate transformation is made using the following relations: $t=x_0$, $r=\exp{(x_1)}$, $\theta=\pi x_2$, and $\phi=x_3$. See Appendix B in \citet{Chatterjee:19} for the detailed coordinate conversion.}}.  This conversion leads to a logarithmic spacing in $r$ such that the cells have a higher resolution for smaller values of $r$.  The spatial resolution near the event horizon is $\Delta r \approx 0.01\,r_{\rm g}$ for the model with the highest resolution. To prevent the aspect ratio of the cells from becoming too large near the polar singularity, we reduce the resolution in $\phi-$direction gradually towards both poles.  We use outflow boundary conditions for both inner and outer radial boundaries, and reflecting boundary conditions in the $\theta-$direction.  Note that the inner boundary is causally disconnected from the flow,
as it is located within the event horizon.

It is common for GRMHD simulations to crash if either the density or the internal energy become very low, particularly in the funnel region along the polar axes or near the outer radial boundaries.  To avoid this, we apply numerical floors for the density and the internal energy (see Appendix B3 of \citealt{Ressler:17} for more detailed discussions): a minimum rest mess density, $\rho_{\rm fl} = {\rm max} \left[ b^2/20,\, u_g / 150,\, 10^{-6} \left( r/r_{\rm g} \right)^{-2} \right]$, and a minimum internal energy density, $u_{g, \rm fl} = {\rm max} \left[ b^2/750,\, 10^{-7} \left( r/r_{\rm g} \right)^{-2 \Gamma} \right]$, where $b$ and $u_g$ are the co-moving magnetic field strength and the internal energy density, respectively. We normalise the mass density such that the maximum density is $\rho_{\rm max} = 1.$

\subsection{Simulation Models}
\label{subsec:models}

\begin{table*}
	\caption{Simulation setup parameters. All models are initialised with $a_{*}=0.9375$, $r_{\rm in}=6~r_{\rm g}$, and $r_{\rm max}=12~r_{\rm g}$.}
    \label{tab:models}
	\begin{center}
	\begin{tabularx}{0.675 \textwidth}{l c c c c c}
		\hline\hline 
%		Model & $a_{\star}$ & $r_{\rm in}\,(r_{\rm g})$ & $r_{\rm max}\, (r_{\rm g})$ \TopTabspace\BotTabspace \\
%		\hline \\
%		All & 0.9375 & 6 & 12 \\
%		\hline \hline \\
		Model Name & Cooling & $\rho_{\rm scale}^a$ & $T_i/T_e$ & $\langle \dot{M} \rangle^b$ ($10^{-8}\,\MsunYr$)  & Resolution  \TopTabspace\BotTabspace \\
%		Name	 &         &    &                  & ($10^{-8}\,\MsunYr$) & \\
	\hline
	    \texttt{C3D01RM} & on    & $1\times10^{-17}$ & 3 & $0.08 \pm 0.01$ & $256\times160\times160$ \TopTabspace \\
		\texttt{C3D1RL}	& on	& $1\times10^{-16}$ & 3 & $0.23 \pm 0.07$ & $128\times64\times64$ \\
		\texttt{C3D1RM}	& on	& $1\times10^{-16}$ & 3 & $1.15 \pm 0.31$ & $256\times160\times160$ \\
		\texttt{C3D1RH}	& on	& $1\times10^{-16}$ & 3 & $1.27 \pm 0.17$ & $648\times384\times384$ \\
		\texttt{C3D1RMFT20}	& on & $1\times10^{-16}$ & 20 & $1.08 \pm 0.20$ & $256\times160\times160$ \\
		\texttt{C3D1RMRh20}	& on & $1\times10^{-16}$ & $R_l$=1\,, $R_h$=20 & $1.13 \pm 0.11$ & $256\times160\times160$ \\
		\texttt{C3D10RM}	& on	& $1\times10^{-15}$ & 3 & $8.22 \pm 1.97$ & $256\times160\times160$   \\
		\texttt{C3D100RM}	& on	& $1\times10^{-14}$ & 3 & $77.82 \pm 14.18$ & $256\times160\times160$   \\
		\texttt{NC3RM}	& off	& --			& --	& --  & $256\times160\times160$   \\
		\texttt{NC2RH}	& off	& --			& --	& --  & $648\times384\times1^{c}$  \BotTabspace \\
	\hline
	\end{tabularx}
	\end{center}
	\begin{itemize}
		\item[$^a$]conversion factor for the mass density from code units to c.g.s units.
		\item[$^b$]mass accretion rate at the event horizon, which is averaged over 3000 -- 8000 $t_{\rm g}$.
		\item[$^c$]axisymmetric 2.5D run for the purpose of comparison.
	\end{itemize}
\end{table*}

We perform a set of GRMHD simulations, in which the magnetised gas is accreting onto a supermassive and spinning BH. All simulations are initialised with a steady-state hydrostatic torus around a rapidly spinning Kerr BH \citep{Fishbone:76}. We set the spin parameter to $a_\star = 0.9375$ for all models. The size of the initial torus is set by the inner edge, $r_{\rm in} = 6\,r_{\rm g}$, and the radius of the pressure maximum, $r_{\rm max} = 12\,r_{\rm g}$. We adopt an ideal gas equation of state,
\begin{equation}
   P_g = \left( \Gamma - 1 \right)\,u_g \,,
\end{equation}
where $P_g$ and $u_g$ are the gas pressure and internal energy, respectively.  We set the adiabatic index to $\Gamma = 5/3$, which assumes the dominance of a non-relativistic plasma in the accretion flow.

As an initial magnetic configuration, we adopt a single loop of weak magnetic field, which is computed from the magnetic vector potential,
\begin{eqnarray}
	A_{\phi} &\propto& \left\{ 
	  \begin{array}{ll}
		  \rho - 0.2,   & {\rm if}~~ \rho > 0.2\,, \\
		  0\,,            & {\rm otherwise}.  
	  \end{array} \right.
\end{eqnarray}
The centre of the loop is at the density maximum, and the loop is fully contained within the initial torus. The initial magnetic field is normalised such that $\beta_{\rm mag}= P_g/P_{\rm B} \geq 100$. This normalisation ensures that the magnetic pressure is subdominant compared to the gas pressure.

\subsection{Radiative Cooling}
\label{subsec:cool}

%It is widely accepted that the mass accretion rate of \sgra{} should be less than $\sim 10^{-7}\,
%\MsunYr$ by constraints on the Faraday rotation \citep{Bower:03,Marrone:07}, below which the
%radiative cooling loss may not be critical. However, cooling becomes more important with higher
%accretion rate as it reduces the plasma pressure and decrease the scale height of the disk
%\citep{Fragile:09}. 
We take into account the radiative cooling self-consistently in our calculation of the gas temperature, by including the effects of bremsstrahlung, synchrotron, and the inverse Compton losses. We adopt the equations of \citet{Esin:96} for computing the radiative cooling losses. These formulae have been implemented and tested in previous numerical studies of \sgra{} \citep{Fragile:09,Dibi:12,Straub:12,Drappeau:13}.

The total cooling rate for an optically thin gas is computed from the cooling function, 
\begin{equation}\label{eq:coolratethin}
	q_{\rm thin}^- = \eta_{\rm br,C}\, q_{\rm br}^{-} + \eta_{\rm s,C}\, q_{\rm s}^{-}\,,
\end{equation}
where $q_{\rm br}^{-}$ and $q_{\rm s}^{-}$ are the bremsstrahlung and synchrotron cooling rates, respectively, and $\eta_{\rm br,C}$ and $\eta_{\rm s,C}$ are the Compton enhancement factors, which are the average energy gain of the photon in an assumption of single scattering \citep{Esin:96}. We note that the Compton enhancement of the bremsstrahlung is negligible as synchrotron is dominant at the temperature where the Comptonization becomes important.

While the whole system is generally optically thin, we use the following generalised cooling formula, from \citet{Narayan:95b} and \citet{Esin:96}, to reproduce the equilibrium solution corresponding to the optically thick disk \citep{Shakura:73}:
\begin{equation}\label{eq:coolrate}
	q^- = \frac{4\,\sigma_{\rm T}\,T_e^4 / H_{\rm T}}{1.5\tau + \sqrt{3} + \tau_{\rm abs}^{-1} }\,,
\end{equation}
where $\sigma_{\rm T}$ and $T_e$ are the Thomson cross-section and the electron temperature, respectively, and the local temperature scale height $H_{\rm T}$ is computed from
\begin{equation}
	H_{\rm T} = \frac{T_e^4}{|\nabla (T_e^4)|}\,.
\end{equation}
The scale heights are locally calculated such that $T_e^4$ drops off by a factor of $1/e$, which was adopted in \citet{Fragile:09} as a suitable and robust treatment in multi-dimensional simulations.

The total optical depth of the disk is calculated by $\tau=\tau_{\rm es} + \tau_{\rm abs}$, where $\tau_{\rm es}=2\,n_e\,\sigma_T H_{\rm T}$ is the Thomson optical depth in the vertical direction and $\tau_{\rm abs}$ is the optical depth for absorption, which is expressed as
\begin{equation}
	\tau_{\rm abs} = H_{\rm T}\, \frac{q_{\rm thin}^-}{4\sigma_{T}\,T_e^4}\,.
\end{equation}
For a small optical depth, \eq{eq:coolrate} reduces to \eq{eq:coolratethin}, while, in the optically thick limit ($\tau \gg 1$), it gives $q^-=8\sigma_{T}\,T_e^4/3H_{\rm T}\tau$, which is the appropriate black body limit. Therefore, the formula provides an approximate interpolation between the optically thin and thick limits. 

%\subsubsection{Bremsstrahlung}
%\label{sub2sec:brem}

At low temperatures ($T_{e} \lesssim 6\times10^{9}$ K) or the outer torus regions, the emission is dominated by bremsstrahlung \citep{Straub:12}. The bremsstrahlung cooling rate is computed by the interactions of pairs among electrons ($e^-)$, positrons ($e^+)$ and ions ($i$). Since the cooling processes of $e^-\,i$ and $e^+\,i$ are identical, and the same is true for $e^-\,e^-$ and $e^+\,e^+$, the cooling rate can be written as,
\begin{equation}
	q_{\rm br}^- = q_{\rm ei}^- + q_{\rm ee}^- + q_{\pm}^-\,,
\end{equation}
where $q_{\rm ei}^-$, $q_{\rm ee}^-$ and $q_{\pm}^-$ are the radiative cooling through electron(positron)-ion ($e^\pm\, i$), electron(positron)-electron(positron) ($e^\pm \,e^\pm$) and positron-electron ($e^+\, e^-$) interactions, respectively \citep{Esin:96}.

However, for most regions of inner hot accretion flows, the synchrotron emission dominates the losses as the electrons are relativistic due to the high electron temperature. The synchrotron cooling occurs through both optically thick and thin emission: below some critical frequency $\nu_c$, the emission is completely self-absorbed, and thus the volume emissivity can be approximated by the Rayleigh-Jeans black body emission. For frequencies above $\nu_c$, the emission is optically thin. The synchrotron cooling rate can be written as,
\begin{equation}\label{eq:syn}
	q_{\rm s}^- = \frac{2\pi\,k_B\,T_e}{H_{\rm T} c^2} \int_0^{\nu_c} \nu^2\,{\rm d}\nu 
				  + \int_{\nu_c}^\infty \epsilon_{\rm s} (\nu)\, {\rm d}\nu\,,
\end{equation}
where $k_B$ and $c$ are the Boltzmann constant and the speed of light, respectively, and the synchrotron emissivity $\epsilon_{\rm s} (\nu)$ is calculated as \citep[see][]{Pacholczyk:70},
\begin{equation}\label{eq:emis}
	\epsilon_{\rm s}(\nu) = \frac{e^2}{c \sqrt{3}} 
						    \frac{4\pi\,\nu \left( n_e + n_+ \right)}{K_2 (1/\Theta_e)}\,I'(x_M)\,,
\end{equation}
where $K_2$ is the modified Bessel function of the second kind,
\begin{equation}
	x_M\equiv\frac{2\,\nu}{3\nu_0\, \Theta_e^2}\,, \qquad v_0 \equiv \frac{e\,B}{2\pi\, m_e\, c}\,,
\end{equation}
and $\Theta_e \equiv k_B\,T_e / m_e c^2$ is the dimensionless electron temperature. The dimensionless spectrum $I'(x_M)$, which is averaged over the angle between the velocity vector of the electron and the direction of the local magnetic field, is fitted by the function \citep{Mahadevan:96},
\begin{equation}
	I'(x_M) = \frac{4.0505}{x_M^{1/6}} \left( 1 + \frac{0.4}{x_M^{1/4}} + \frac{0.5316}{x_M^{1/2}} \right)\,
			  \exp{\left( -1.8899\,x_M^{1/3} \right)}\,.
\end{equation}
\citet{Fragile:09} found that the Bessel function $K_2$ in \eq{eq:emis} causes errors for the low-temperature flows ($T_e < 10^8$ K) due to the mismatch of the normalisation factor between the Bessel function and the spectrum $I'(x_M)$. Following their suggested modification, we replace $K_2(1/\Theta_e)$ by $2\,\Theta_e^2$, thereby assuming the same high-temperature limit.  

%\cmtdy{add high temperature limit}

We numerically compute $\nu_c$ in \eq{eq:syn} by equating the optically thin and thick volume emissivities at $\nu_c$,
\begin{equation}
	\epsilon_{\rm s}(\nu_c) = 
                            \frac{e^2}{c \sqrt{3}} 
						    \frac{4\pi\,\nu_c \left( n_e + n_+ \right)}{K_2 (1/\Theta_e)}\,I'(x_M)
	                       = \frac{2\pi k_{B}\,T_e}{H_{\rm T}\,c^2} \nu_c^2\,.
\end{equation}

% Shcherbakov+12: non-thermal electrons (Mahadevan 1998;Ozel¨ et al. 2000; Yuan et al. 2004).

\section{Results}
\label{sec:results}

%We carry out the full 3D GRMHD high-resolution simulations with  radiative cooling
%for studying the dynamics of the accreting plasma and the resulting spectra around \sgra{}. Such
%simulations are computationally expensive, which limits the physical parameter space that we can
%explore. 

We initialise our fiducial model by following the ``best-bet'' model, that is widely agreed by previous 2.5D parameter surveys \citep[e.g.,][]{Moscibrodzka:09,Dibi:12,Drappeau:13}, which have $T_i/T_e = 3$ and $a_\star = 0.9375$ (see Table~\ref{tab:models}).

\subsection{General Evolution}

\begin{figure}
	\centering
	\includegraphics[width=\columnwidth]{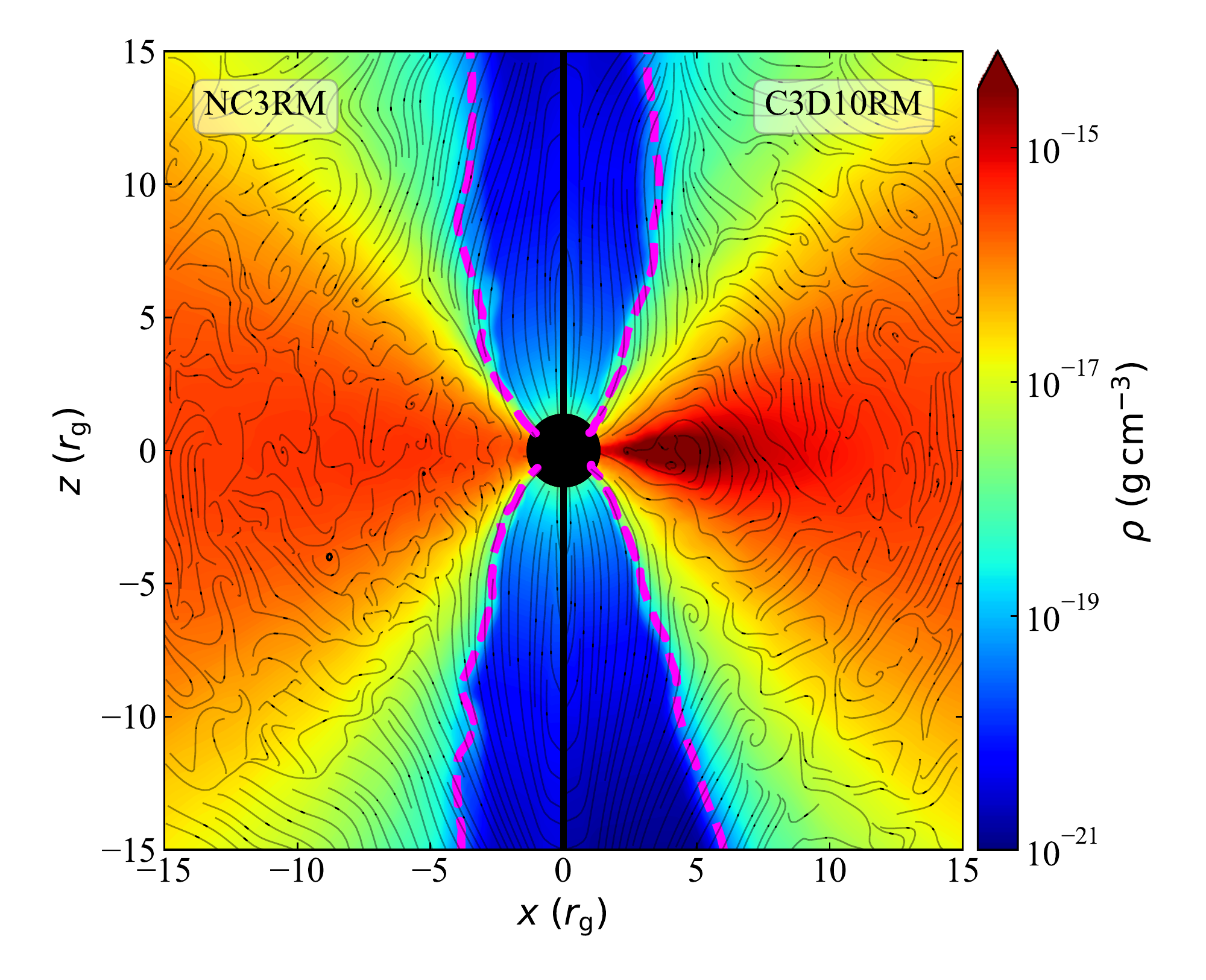}
	\caption{Comparison of the time-averaged density contour map between the \texttt{NC3RM} (left, non-cooled) and \texttt{C3D10RM} (right, cooled) simulations.  The non-cooled model is scaled by multiplying the same density unit (see $\rho_{\rm scale}$ in Table~\ref{tab:models}) for the corresponding cooled model. Given the density unit, the average mass accretion rates are $8.24 \times10^{-8}\,\MsunYr$ and $8.22\times10^{-8}\,\MsunYr$ for the (scaled) non-cooled and cooled models, respectively. The light black contours represent the magnetic field, and the dashed magenta lines denote the jet boundary, defined as where the magnetisation parameter is $\sigma=1$. All variables are averaged over $5000~t_{\rm g}$ -- $6000~t_{\rm g}$.}
	\label{fig:dcmap}
\end{figure}

Our simulations start with the initial torus in hydrostatic equilibrium \citep{Fishbone:76}. As the turbulence triggered by tangled magnetic fields transports angular momentum outward, the gas flows towards the central BH generating a thick disk, akin to a RIAF. The cooled models require a specific density unit to be pre-set in order to achieve the designated mass accretion rate in the simulations (see Table~\ref{tab:models}). However, the non-cooled models are scale-free and therefore scaled by the corresponding density unit (in GRRT post-processing) to enable comparison with the cooled models.  
%While the non-cooled model is  simulation is not scaled by density unit (i.e., scale-free), the cooled simulation are required to set the specific density unit to target the designated mass accretion rate (see Table~\ref{tab:models}).

\fig{fig:dcmap} shows the density contour map overlaid with the magnetic field structure, at which the data is averaged for 5000 -- 6000 $t_{\rm g}$\footnote{The orbital time scale at the pressure maximum, $r_{\rm max} = 12\,r_{\rm g}$, is $t_{\rm orb} \sim 260\,t_{\rm g}$ in our simulations, where $t_{\rm g}\sim20~{\rm s}$ for \sgra{}.}. The overall evolution of the accreting hot accretion flow is similar between the non-cooled and cooled runs when the mass accretion rate is smaller than $10^{-8}\,\MsunYr$.  However, the effect of cooling becomes increasingly important and shows visible differences in model \texttt{C3D10RM}, where the target mass accretion rate is $10^{-7}\,\MsunYr$ (the estimated value is $8.22\times10^{-8}\, \MsunYr$): when radiative cooling is on, the density increases significantly in the mid-plane and the magnetic field within the disk is less turbulent. This is because cooling reduces the gas pressure and the corresponding scale height of the accretion flow, thus increasing the dominance of magnetic fields: the plasma beta, $\beta_{\rm mag}\equiv P_g/P_{\rm B}$, decreases due to the reduced gas pressure and the compressed volume. Such highly magnetised plasma tends to be stable against the MRI, and thus the magnetic field within the disk becomes less turbulent.

%Given the low accretion rate, the overall evolution of the accreting
%hot accretion flow is similar between the non-cooled and cooled runs, which is not surprising since
%effect of radiative cooling is not significant in this low accretion regime: the enhanced density at
%the mid-plane overlapped with a tangled, turbulent magnetic fields, and the apparent bipolar jets or
%``funnel'' over the poles of BH (indicated by magenta dashed-lines; where the magnetization
%parameter $\sigma=1$) along the ordered magnetic field lines.

\begin{figure}
	\centering
	\includegraphics[width=\columnwidth]{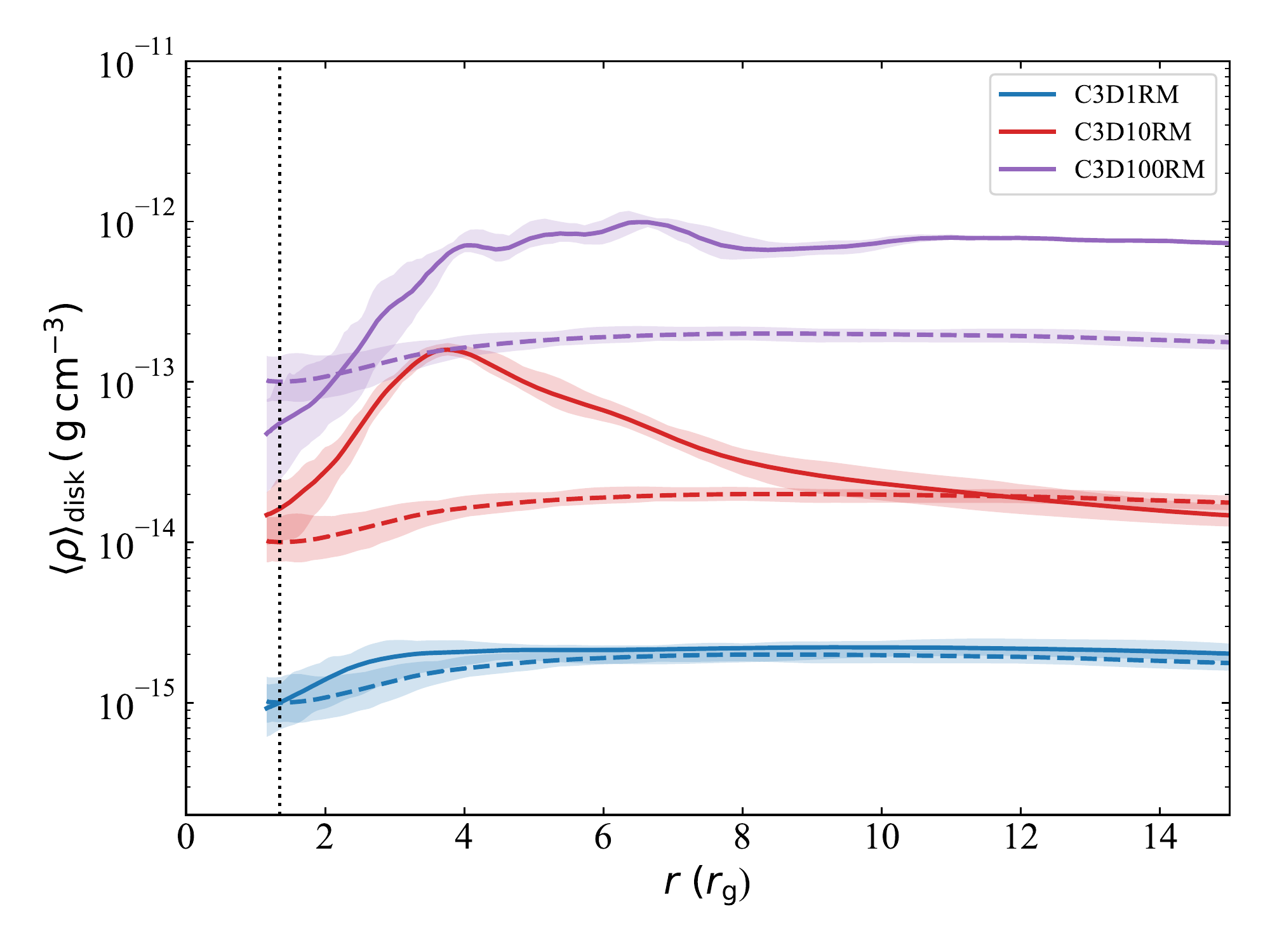}
	\caption{The averaged density profile along the disk over the time interval $6000~t_{\rm g}$ -- $8000~t_{\rm g}$ (see Eq.~\ref{rhoaverage}). The solid lines represent the result from models \texttt{C3D1RM}, \texttt{C3D10RM}, and \texttt{C3D100RM}, and the dashed lines are the profile from the non-cooled model \texttt{NC3RM} with re-scaling to each of the cooled models. The vertical dotted line indicates the location of the event horizon, $r_H = r_{\rm g} \left( 1+\sqrt{1-a_\star^2} \right) = 1.35\,r_{\rm g}$.}
	\label{fig:dcomp}
\end{figure}

Radiative cooling enhances the mid-plane density for two main reasons: at first, it is clear that cooling reduces the gas pressure as thermal energy is radiated away. Secondly, the relatively ordered magnetic field impedes angular momentum transport through MRI, so accretion slows down and piles up where the MRI is less efficient. \fig{fig:dcomp} shows the averaged density profile along the disk over the time interval between $6000~t_{\rm g}$ -- $8000~t_{\rm g}$ by the formula:
\begin{equation}\label{rhoaverage}
	\langle \rho \rangle (r) := \frac{\int^{2\pi}_0 \int^{\pi}_{0} \rho' \,\sqrt{-g}\, {\rm d}\theta\, {\rm d}\phi}
	                   {\int^{2\pi}_0 \int^{\pi}_{0} \sqrt{-g}\, {\rm d}\theta\, {\rm d}\phi}\,,
\end{equation}
where $\rho'$ is the time-averaged density. In the figure, the density enhancement is apparent as a consequence of cooling. For models with higher cooling, the relative increase in the density compared to non-cooled runs is larger and occurs over a broader range in radius. More importantly, the peak of the averaged density is located at larger distances and with stronger cooling, which is not surprising because the angular momentum is more difficult to transport outward if cooling is strong. The location of the peak density also affects the mass accretion rate: if it is close to the vicinity of the BH, the accretion rate can increase slightly due to the increased density near the event horizon (see \texttt{C3D10RM} model in \fig{fig:dcomp}). Its effects on the resulting spectra may not be trivial since a large fraction of synchrotron radiation is produced near the BH, as discussed in \S~\ref{subsubsec:spec}.

%Radiative cooling, however, enhances the density at the vicinity of the BH, where $r < 5\,r_{g}$ (see
%Figure~\ref{fig:dcomp}). In the figure, we see that for the models with the cooled simulations, the
%disk-averaged density at the event horizon is a factor of two larger, compared to the non-cooled
%simulations.  Although this density enhancement does not affect the global evolution, its effect on
%the resulting spectra is not trivial since a large fraction of synchrotron radiation is produce near
%the BH as will be discussed in \S~\ref{subsec:spec}.  The disk-averaged density profile in the
%figure is computed over the time interval between 5000 -- 6000 $t_g$ by the formula:

\begin{figure*} 
	\centering
	\includegraphics[width=0.77\textwidth]{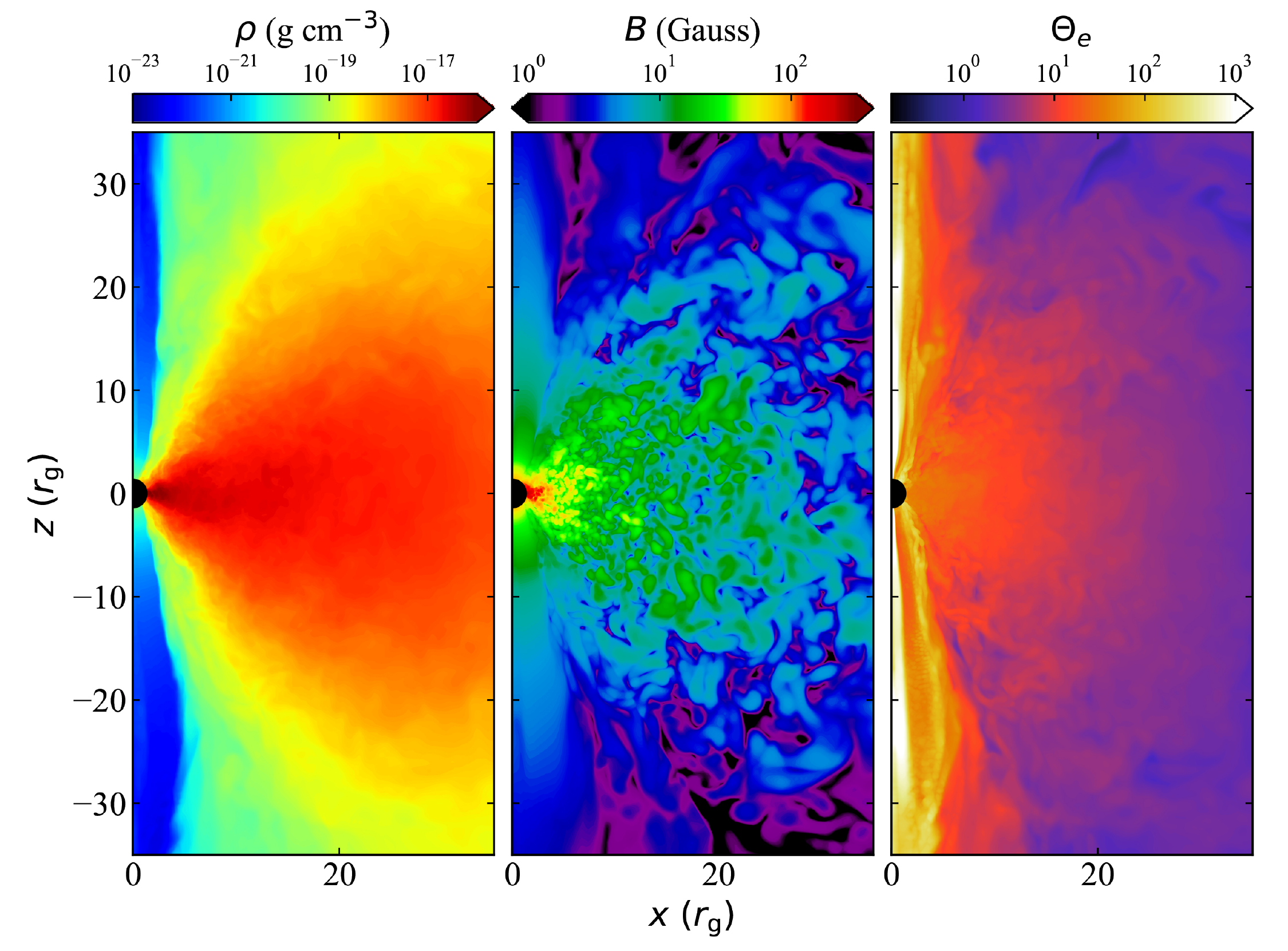}
	\caption{Colour contour maps of the mass density, the magnetic field strength, and the electron temperature for the model \texttt{C3D1RH}. The figure presents a single time slice at 5000 $t_{\rm g}$.}
	\label{fig:phycont}
\end{figure*}
\begin{figure*} 
	\centering
	\includegraphics[width=0.77\textwidth]{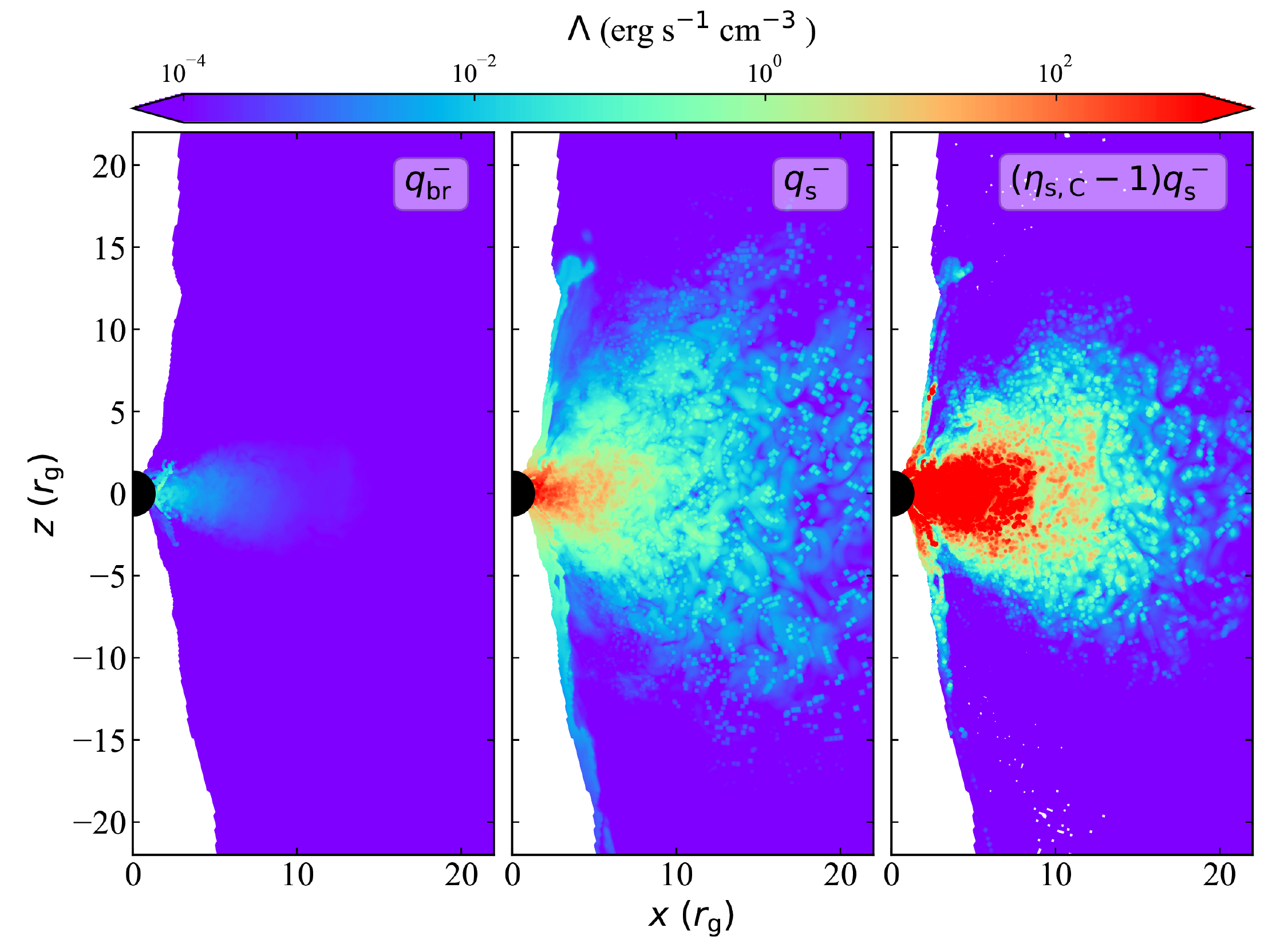}
	\caption{Colour contour maps of the radiative cooling rates: bremsstrahlung (left), synchrotron (middle), inverse Compton scattering (right) for the model \texttt{C3D1RH}. The figure presents a single time slice at 5000 $t_{\rm g}$. Radiative cooling is calculated only when $B^2/\rho < 1$  (see the white area, where the cooling process is off).}
	\label{fig:radcont}
\end{figure*}

Radiation processes occur predominantly in hot accretion flows. To obtain insight into the physical properties of the flows, we show contour maps slices of density, $B^2$, and electron temperature at a single timestep ($t=5000\, t_{\rm g}$) for the highest resolution model (\fig{fig:phycont}). For the hot accreting plasma, it is evident that the typical density is $\sim 10^{-16} \,\rm g\,cm^{-3}$, which corresponds to $N_e\approx 6\times10^7\,\rm cm^{-3}$ in fully ionized plasma, where $N_e$ is the electron number density. The typical strength of magnetic fields is B$\simeq~30\,{\rm G}$ near the event horizon, which decreases with increasing distance from the BH. 
The electron temperature is maximal within the ``funnel'' over the pole, which is up to $\sim 10^{13}$ K, and is $\sim 10^{12}$ K in the mid-plane. Despite the high temperature in the funnel, it typically produces negligible emission as a consequence of the extremely low densities in this region. Note that since we assume a relativistic thermal Maxwell-J\"uttner  distribution for the radiative processes, the question still remains as to which fraction of non-thermal electrons can be generated  within the plasma by steepening of MHD waves or inducing shocks/turbulence through mechanisms akin to magnetic reconnection, and thus how it contributes to the radio emission \citep{Yuan:03,Davelaar:18}.
%\citep[e.g., $\kappa$ - distribution in ][]{Davelaar:18}. 

Given the electron temperature range of $10^{11}$ -- $10^{12}$ K in the accretion flow, as seen in \fig{fig:radcont}, bremsstrahlung cooling is relatively weak. Note that in our simulations the Comptonization of bremsstrahlung is neglected as it is never of importance compared to synchrotron emission over the temperature range of interest. On the other hand, synchrotron cooling with Compton enhancement is dominant in the mid-plane near the BH: optically thin synchrotron radiation at $r\lesssim 5\,r_{\rm g}$ is responsible for the sub-mm peak in the spectral energy distributions (SEDs), which lies within $10^{11}~{\rm Hz} < \nu < 10^{14}~{\rm Hz}$ (see \fig{fig:grmonty_spec}).   Inverse-Compton scattering of synchrotron photons is active at $r\lesssim 8\,r_{\rm g}$. The mean electron temperature in the accretion flow at $r\lesssim 8\,r_{\rm g}$ is $\langle T_{e} \rangle \approx 3\times 10^{11}$ K, and therefore the average increase of energy in a single scattering can be approximated as $A = 1 + 4\Theta_e + 16\Theta_e^{2} \approx 4\times10^4$ \citep[see][]{Esin:96}.  As a result, the frequencies of these scattered photons are shifted to the range $0.02~{\rm keV}$ -- $20~{\rm keV}$, which is consistent with the observed emission in X-rays ($0.5~{\rm keV}$ -- $8~{\rm keV}$) \citep{Baganoff:01,Baganoff:03}. The quiescent X-ray emission of \sgra{} is extended, with an intrinsic size of $\sim 1.4''$ \citep{Baganoff:03}, which is coincident with the Bondi accretion radius calculated from the measured BH mass and ambient temperature \citep{Yuan:03}. It is known that $\sim$90\% of the total X-ray emission originates from the outer part of the disk \citep{Neilsen:13}, and is dominated by bremsstrahlung, which is beyond the scope of this work: we calculate the emission within $r=20~r_{\rm g}$, where the X-ray emission is predominantly produced by the synchrotron self-Compton (SSC) process. We will further discuss the spectral properties of the X-ray emission in \S~\ref{subsec:radiation}.

\subsection{Mass accretion rates}

\begin{figure} 
	\centering
	\includegraphics[width=\columnwidth]{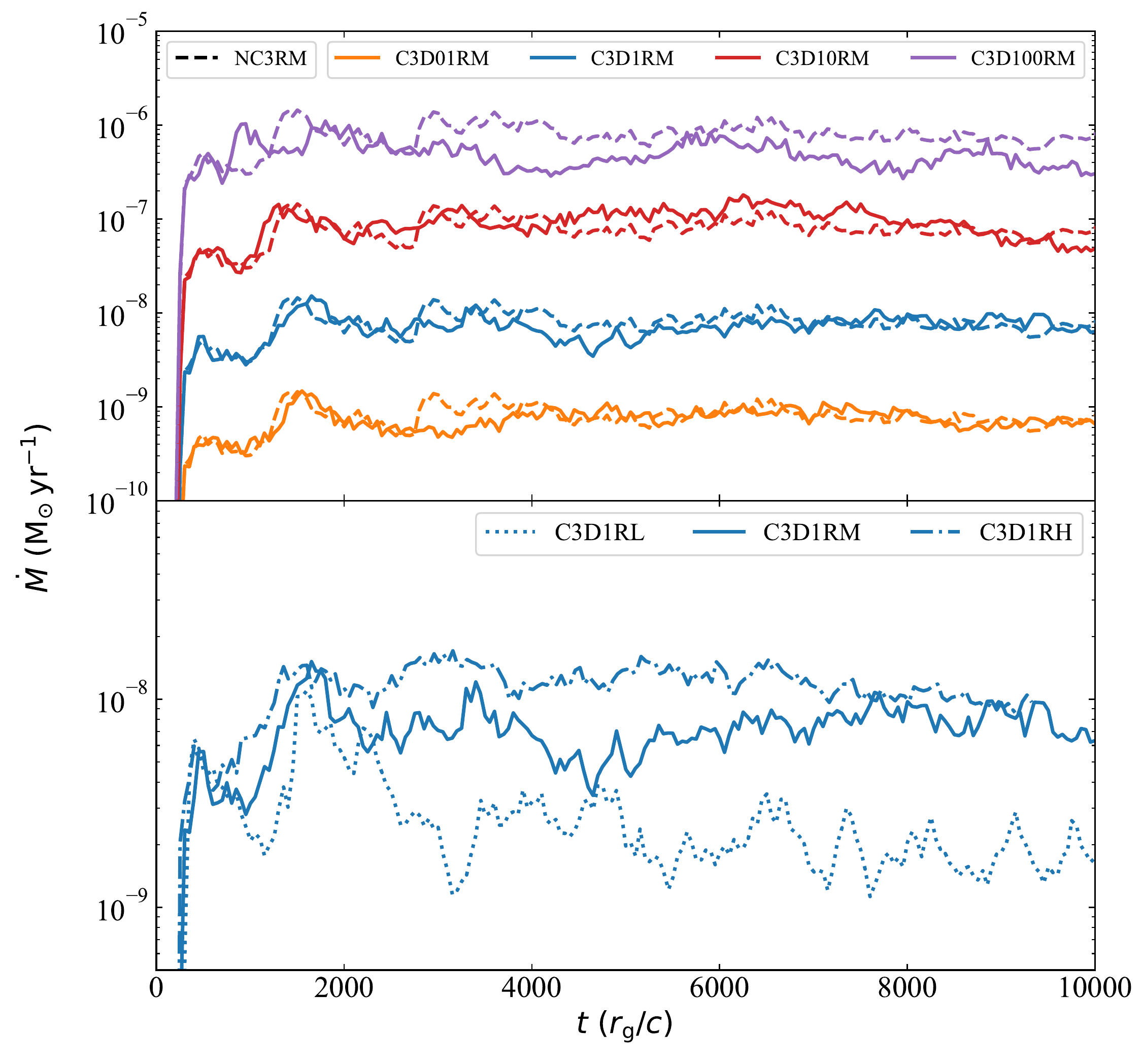}
	\caption{The mass accretion rate at the event horizon as a function of time.  Upper panel: The solid and dashed lines represent the results from models with radiative cooling, and models without radiative cooling (\texttt{NC3RM}), respectively. The accretion rate from the \texttt{NC3RM} model is scaled by the density unit that corresponds to the each cooled counterpart.  Lower panel: The accretion rate from the simulations at the different resolutions: \texttt{C3D1RH} (high, dot-dashed), \texttt{C3D1RM} (intermediate, solid), \texttt{C3D1RL} (low, dotted).} \label{fig:massaccr}
\end{figure}

The mass accretion rate is a critical factor in determining the radiation fluxes. In previous works which omit radiative cooling, the simulations are scale-free and must therefore be scaled with an arbitrary density unit during the GRRT post-processing to achieve the desired accretion rate. The scaled variables are used to calculate synthetic spectra which match with observations \citep[e.g.,][]{Dexter:10,Shcherbakov:12,Moscibrodzka:14}. Our radiatively-cooled simulations, however, are not scale-free: the calculation of cooling rates requires specifying variables in physical units. 

\fig{fig:massaccr} shows the mass accretion rates over time until $t=10,000\,t_{\rm g}$, which is calculated by
\begin{equation}
	\dot{M}(r) = \int_0^{2\pi}\int_0^{\pi}\,\rho \,u^{r}\sqrt{-g}\,{\rm d}\theta\,{\rm d}\phi\,,
\end{equation}
where $u^r$ is the radial component of the 4-velocity. As seen in the figure, the mass accretion rate converges after 3000 $t_{\rm g}$, which corresponds to $\sim$ 10 orbital time scales at the pressure maximum. This is the case for all models except the low resolution runs (\texttt{C3D1RL}). The convergence of the accretion rate allows us to study the statistical properties over a longer time period. In contrast to the 3D runs, it is known that previous 2.5D simulations fail to reach the steady-state of the accretion as a consequence of the anti-dynamo theorem \citep{Hide:82}. We further compare the results between 2.5D and 3D runs in \S~\ref{subsec:comp2D3D}.

In the lower panel of \fig{fig:massaccr}, we show how the angular momentum transport through the turbulence of the accretion flow can be affected by resolution effects. To examine if the MRI is resolved properly, we calculate the ``MRI quality factors'' (i.e., Q-factors), which are defined as the number of cells available for resolving the fastest-growing MRI mode in each direction. The Q-factor of the lowest resolution case (\texttt{C3D1RL}) is $\sim$ 3, which is below the nominal Q value of 10--20 for capturing the saturation level of the MRI \citep{Hawley:11}.  Thus, it is obvious that the mass accretion rate in the model \texttt{C3D1RL} drops significantly after 2000 $r_{\rm g}$ since the low-resolution run fails to resolve MRI-driven turbulence. However, for the intermediate (\texttt{C3D1RM}) and highest (\texttt{C3D1RH}) resolution cases, the Q-factors are 12 and 20, respectively, which are large enough to sustain the MRI-driven turbulence. These Q-factors are indicative of a criterion above which the simulations satisfactorily resolve the MRI, but cannot be used for the analysis of turbulent features in the flow. We will further discuss the disk properties in \S~\ref{subsec:diskproperties}.

We carry out multiple simulations with  different density units to target Sgr A* mass accretion rates (in units of $M_{\odot}\,{\rm yr}^{-1}$) of $10^{-9},\, 10^{-8},\, 10^{-7},\, {\rm and}\, 10^{-6}$. The largest accretion rate amongst these target values is beyond the observed range around \sgra{} ($2\times10^{-9} M_{\odot}\,{\rm yr}^{-1}<\dot{M}<2\times10^{-7} M_{\odot}\,{\rm yr}^{-1}$), but this model is included to compare the results from other simulations with the case of an extremely high accretion rate.  In the upper panel of \fig{fig:massaccr}, the solid and dashed lines represent the mass accretion rates, which are calculated from the cooled and non-cooled models, respectively. The non-cooled model is re-scaled by the same density unit for each cooled model. For the model with strong cooling (\texttt{C3D100RM}), it is clear that the overall accretion rate is smaller by a factor of two, compared to the non-cooled case with the same density unit. However, the models with weak cooling (\texttt{C3D01RM} and \texttt{C3D1RM}) show no significant differences in the accretion rate between the cooled and non-cooled models, which is surprising since the cooled model is expected to lose less angular momentum compared to the non-cooled model. For model \texttt{C3D10RM}, the accretion rate is even slightly higher than in the non-cooled model. The reason for this is the enhanced density in the vicinity of the event horizon playing a role in increasing the accretion rate (see \fig{fig:dcomp}), which compensates for the weak angular momentum transport.

%It is clear that the accretion rate is enhanced in all cooled models (solid
%lines) after $5000\,t_g$, compared to the corresponding non-cooled model (dashed lines), which is
%re-scaled by the same density unit for each cooled model. Given the low accretion ranges, the
%discrepancy is not significant, but the enhancement of the accretion implies that radiative cooling
%makes the accreting flows colder and denser.

\subsection{Disk properties}
\label{subsec:diskproperties}
\begin{figure} 
	\centering
	\includegraphics[width=\columnwidth]{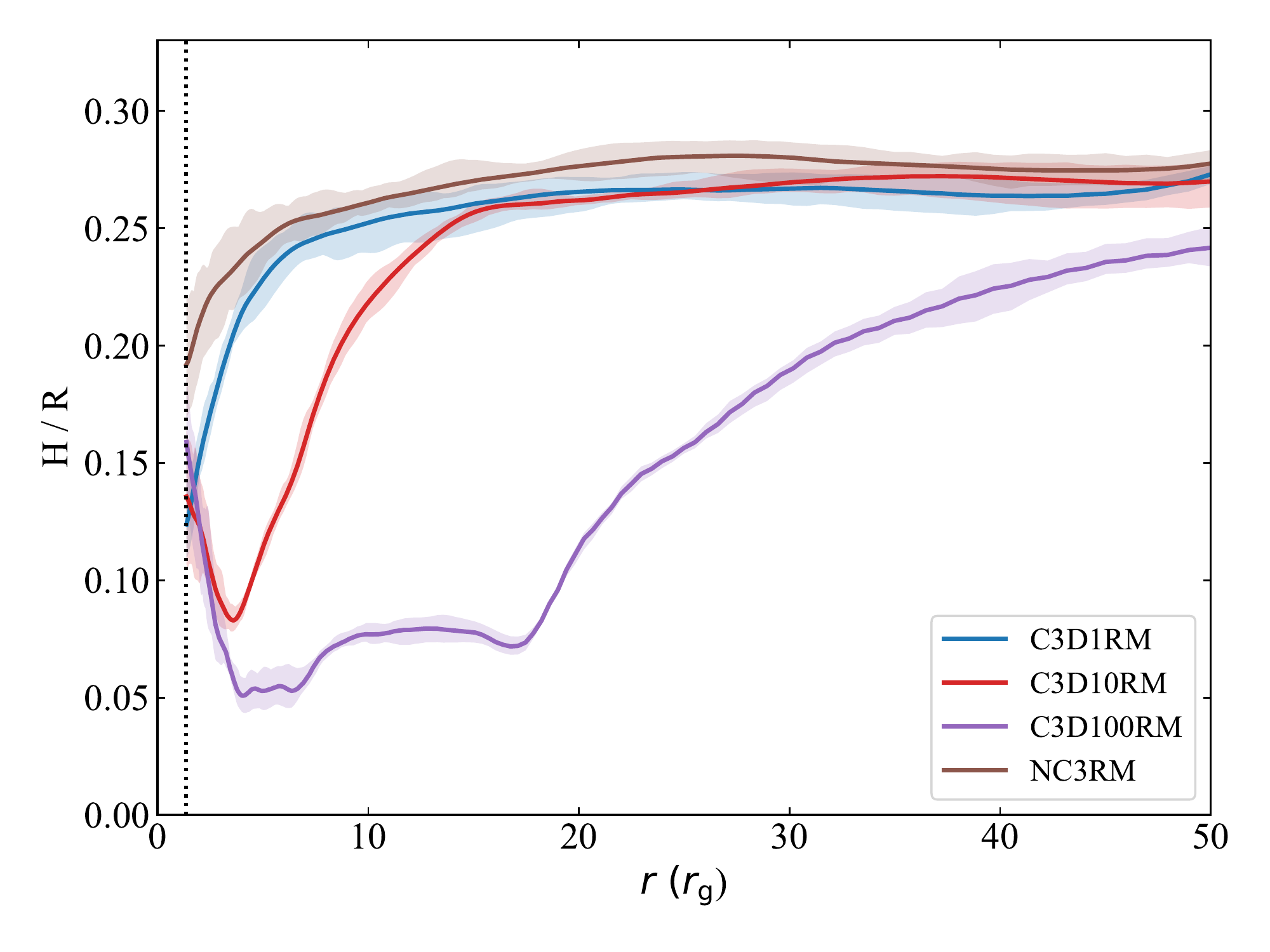}
	\caption{Scale height profiles ($H/R$) according to \eq{eq:HoR} for models with different density scales, and thus different accretion rates. The solid thick lines represent the mean value over a time interval $6000~t_{\rm g}$--$8000~t_{\rm g}$, and the shaded regions represent the variations during this interval.}
	\label{fig:HoR}
\end{figure}

The direct impact of radiative cooling on the accretion flow can be examined through the disk scale height: cooling decreases the gas pressure, and thus renders the disk thinner (see \fig{fig:dcmap}). To examine the scale height quantitatively, we compute the formula in \citet{Noble:10,Porth:19}, which is expressed as 
\begin{equation}\label{eq:HoR}
	[H/R] (r) := \frac{\int_0^{2\pi}\int_0^{\pi} \, \rho\,\sqrt{-g}\,\left| \theta - \pi/2 \right| {\rm d}\theta\, {\rm d}\phi}
	              {\int_0^{2\pi}\int_0^{\pi} \, \rho\,\sqrt{-g}\, {\rm d}\theta\, {\rm d}\phi}\,.
\end{equation}
\fig{fig:HoR} shows a clear trend that as cooling becomes stronger, the disk scale height becomes thinner. While the disk swells up rapidly at $r \lesssim 10\,r_{\rm g}$ in the case of weak cooling (\texttt{C3D1RM}), the increase of $H/R$ is more gradual in the case of stronger cooling (\texttt{C3D10RM}).  Except for the case with extremely strong cooling (\texttt{C3D100RM}), the disk scale heights lie within $H/R=0.24$--$0.28$ for radii within $20\,r_{\rm g} < r < 50\,r_{\rm g}$.
\begin{figure} 
	\centering
	\includegraphics[width=\columnwidth]{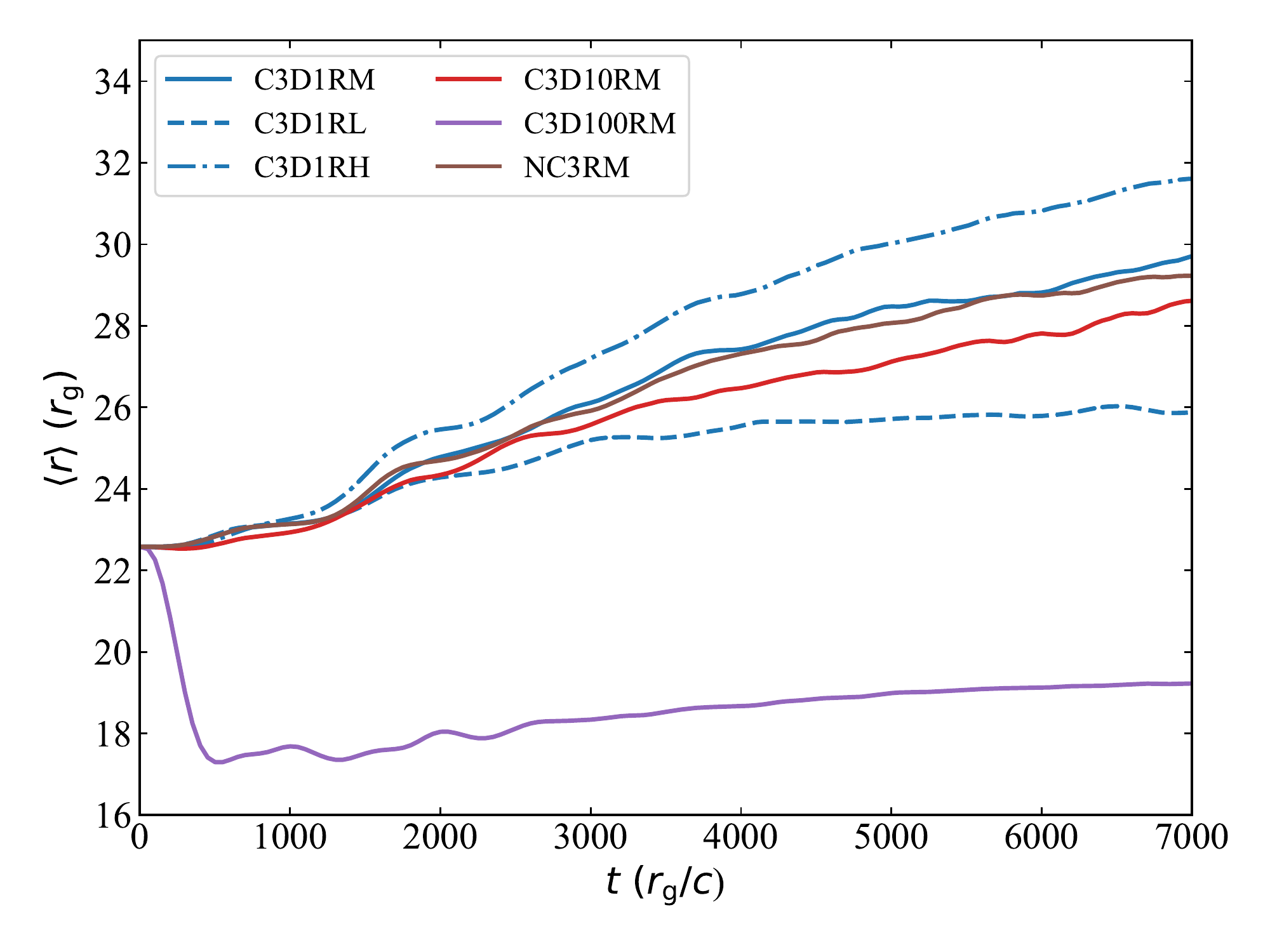}
	\caption{Barycentric radii of the disk (i.e., density-weighted radii, see \eq{eq:rrho}).
	The solid curves represent models with medium resolution but differing density scales,
	and the dot-dashed line and the dashed line represent models that are the same as the model 
	\texttt{C3D1RM}, but with lower (\texttt{C3D1RL}) and higher (\texttt{C3D1RH}) resolutions.}
	\label{fig:rhorad}
\end{figure}
As angular momentum is transported outward by the MRI, the gas flows inward and the disk undergoes viscous spreading outwards. Since radiative cooling reduces the MRI turbulence via the enhanced magnetic field strength (see \fig{fig:dcmap}), less spreading of the disk is expected when the cooling is stronger. For a more quantitative perspective, we compute the rest-frame density-weighted radius, $\langle r_{d} \rangle$ (referring to the formula in \citet{Porth:19}), which is expressed as
\begin{equation}\label{eq:rrho}
	\langle r_d \rangle (t) := \frac{\int_0^{2\pi}\int_0^{\pi}\int_{r_{\rm h}}^{r_{\rm max}} r\,\rho \,\sqrt{-g}\,{\rm d}r\, {\rm d}\theta\, {\rm d}\phi}
								  {\int_0^{2\pi}\int_0^{\pi}\int_{r_{\rm h}}^{r_{\rm max}} \rho \,\sqrt{-g}\,{\rm d}r\, {\rm d}\theta\, {\rm d}\phi}\,,
\end{equation}
where we set the outer radius of integration to $r_{\rm max} = 50\,r_{\rm g}$.  As seen in \fig{fig:rhorad}, the disk spreading is not distinguishable for models \texttt{C3D1RM} and \texttt{NC3RM}, implying that radiative cooling is not strong enough to affect disk spreading for accretion rates up to $\dot{M}=10^{-8}\,\MsunYr$. However, it is apparent that the disk size decreases significantly when the accretion rate is higher than this value. Since current observations of \sgra{} indicate that the accretion rate can reach up to $\dot{M}_{\rm Sgr\, A*,max}=2\times 10^{-7}\, \MsunYr$ \citep{Marrone:07}, the effects of cooling on the dynamics of the accretion flow should be taken into account even within the range of observationally-inferred accretion rates in \sgra{}. The model with the strongest cooling (\texttt{C3D100RM}) shows little spreading over the entire simulation time. Although cooling hinders angular momentum transport as discussed above, the results of \texttt{C3D100RM} may be too dramatic to be considered physically realistic. We found that the MRI Q-factor is reduced to 5-8 for model \texttt{C3D100RM}, as the Alf\'{v}en velocity decreases with increasing density in the mid-plane due to the stronger cooling. This range of the Q-factor lies below the criterion for sufficiently capturing the MRI saturation, thereby the significant changes seen in model \texttt{C3D100RM} may be partially caused by the failure to adequately resolve the MRI. Evidently, the resolution also affects disk spreading: as discussed above, the simulations with lower resolutions cannot capture the MRI sufficiently, resulting in suppressed disk spreading (see the dashed blue line; \texttt{C3D1RL}).

%\cmtdy{working on power spectrum with logarithmic radial grid...}

\section{Discussion}
\label{sec:discuss}

\subsection{Radiative Properties of Sgr~A*}
\label{subsec:radiation}

\subsubsection{Spectral Energy Distribution}
\label{subsubsec:spec}
\begin{figure} 
	\centering
	\includegraphics[width=\columnwidth]{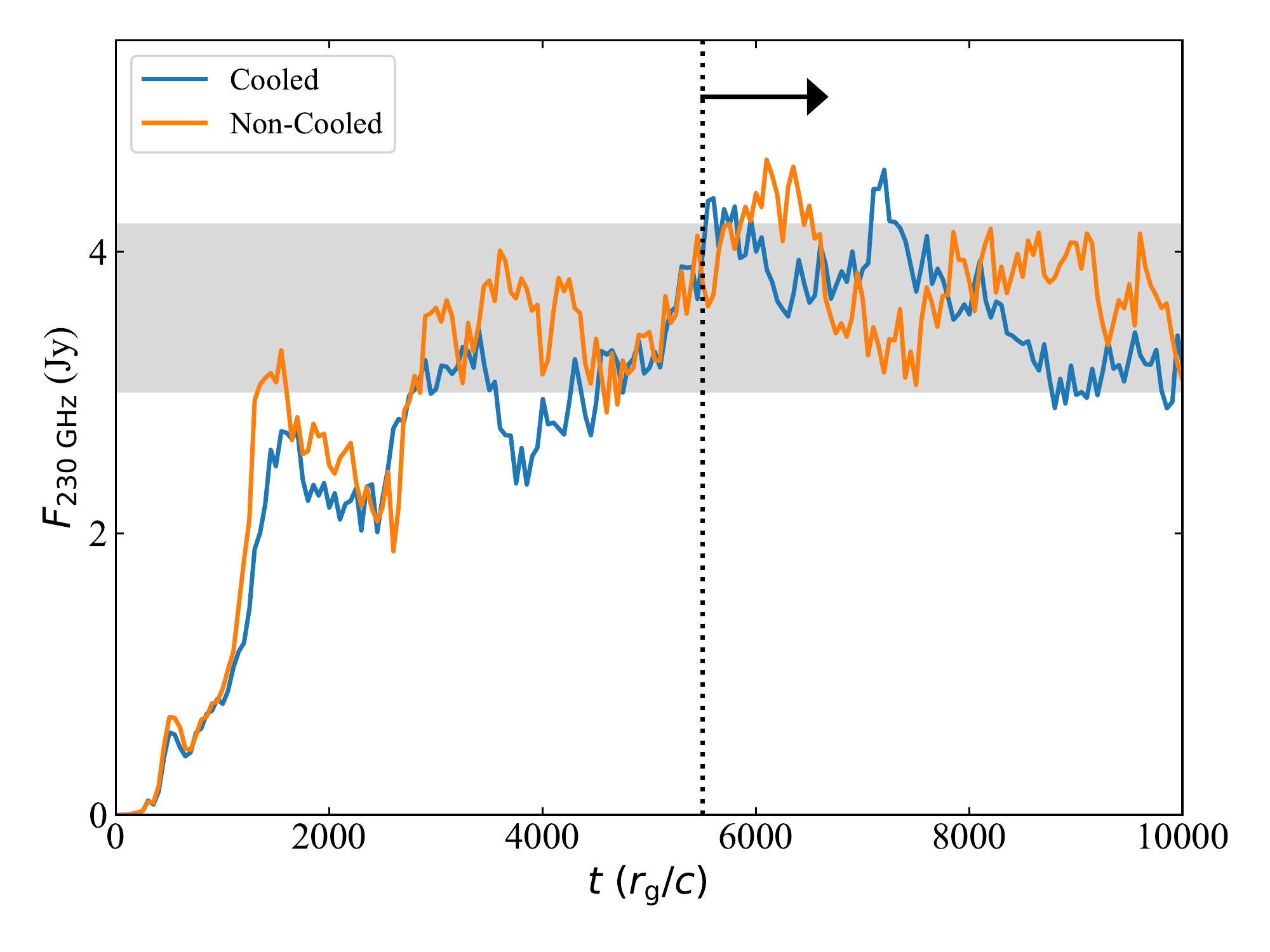}
	\caption{The synchrotron light curves at 230 GHz for the cooled model (\texttt{C3D01RM}; blue color) and the non-cooled model (\texttt{NC3RM}; orange color), which are calculated using \bhoss{}\citep{Younsi:20,Younsi:19_polarizedbhoss}.  The grey shaded region represents the flux range consistent with observations compiled by \citet{Connors:17}. The synthetic spectra (\fig{fig:grmonty_spec}) are calculated after $5500~t_{\rm g}$ (vertical dotted line), where the synchrotron flux lies within the observed range for both the cooled and non-cooled models.}
	\label{fig:bhoss_lightcurve}
\end{figure}
Once we consider radiative cooling, we can no longer scale the GRMHD data to fit the observed flux.  Hence, we choose the best-fit data that produces the flux lying within the observed ranges at 230 GHz ($3~{\rm Jy}$--$4.2~{\rm Jy}$; see observations compiled within \citealt{Connors:17}), but note that this is not a statistical fit.  To compare the results between the cooled data and the non-cooled data, we scale the latter with the same mass density unit for each of their cooled counterparts. As seen in \fig{fig:bhoss_lightcurve}, with our fixed value of $a_{\star}=9375$ and $T_i/T_e=3$, the model \texttt{C3D01RM}, at which the target mass accretion rate is $\dot{M}=10^{-9}\,\MsunYr$, is reasonably consistent with the observations. The overall shapes of the light curves between the cooled (\texttt{C3D01RM}) and non-cooled (\texttt{NC3RM}) models are similar to each other, however the average fluxes in the non-cooled model are slightly higher than in the cooled model. The average fluxes at 230 GHz, which are calculated for the time after 5000 $t_{\rm g}$, are $3.63\pm 0.41$ Jy and $3.8\pm 0.35$ Jy for the cooled and non-cooled models, respectively. 
\begin{figure*} 
	\centering
	\includegraphics[width=\textwidth]{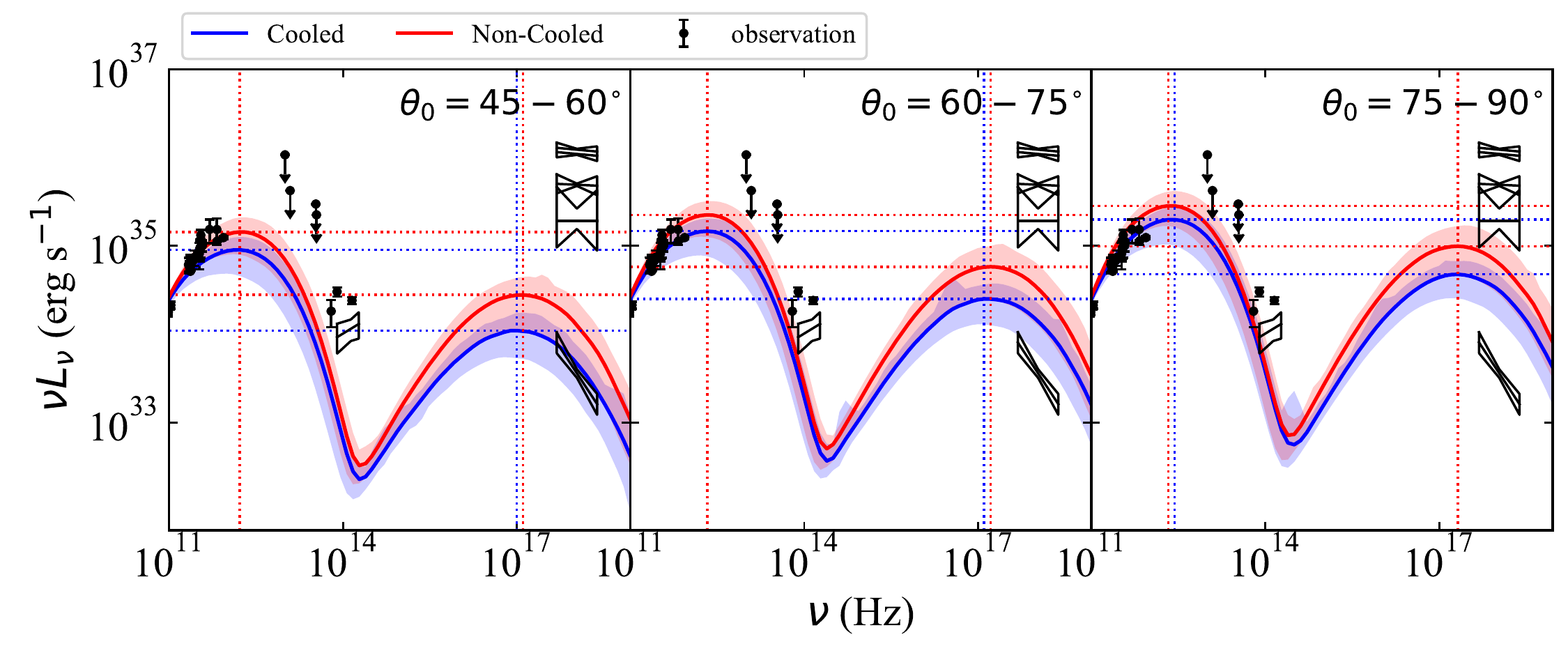}
	\caption{Spectral energy distributions of \sgra{}, which are calculated from the simulation results of \texttt{C3D01RM} (cooled, blue) and \texttt{NC3RM} (non-cooled, red).  The GRMHD data are averaged over the time interval $t_{\rm g}~6000$--$10000~t_{\rm g}$ (see \fig{fig:bhoss_lightcurve} for the light curve within this interval). The solid lines represent the mean value of the spectrum, and the shaded regions represent the variation of the spectrum during the time period. Observational data points are taken from: \citet{Melia:01,Schodel:11} at the upper limit of near-to-mid IR band, \citet{Connors:17} in the sub-mm band, \citet{Bower:19} at terahertz frequencies (233, 678, and 870 GHz), and \citet{Baganoff:01,Baganoff:03} for X-rays (2--10 keV). The X-ray flux in the ray-traced GRMHD data should be below $\sim$10\% of the observed quiescent (lower) X-ray flux since most of the X-rays should be emitted from the outer disk via bremsstrahlung \citep{Wang:13,Neilsen:13}, which is not included in this calculation. The different panels represent the results with different inclination angles: $45^\circ$--$60^\circ$ (left), $60^\circ$--$75^\circ$ (middle), and $75^\circ$--$90^\circ$ (right).}
	\label{fig:grmonty_spec}
\end{figure*}

We calculate the spectra from the GRMHD simulaton data using the Monte Carlo radiative transport code \grmonty \citep{Dolence:09}, which computes synchrotron emission and absorption, and inverse Compton scattering in full general relativity.  \fig{fig:grmonty_spec} shows the SEDs for the cooled model (\texttt{C3D01RM}) and the non-cooled model (\texttt{NC3RM}) with the same density scale. The SEDs have two peaks: the sub-mm peak and the far-UV peak. Thermal synchrotron emission from mildly relativistic electrons is responsible for the sub-mm peak and these same photons are then Compton upscattered to produce the far-UV peak. To check the dependency of the viewing angle, we set the number of $\theta$-bins to 6, within which the fluxes are averaged to represent the values for the range of the inclination angle between the BH spin axis and the observer line-of-sight. In general, the fluxes slightly increase with increasing inclination angle. This is mainly due to the orbiting plasma that is approaching the observer and the emission being more strongly Doppler boosted at higher inclination angles (i.e., close to edge-on).

%\begin{figure} 
%	\centering
%	\includegraphics[width=\columnwidth]{Figs/grmonty_spec_rescale.pdf}
%	\caption{\update{Spectral energy distributions of \sgra{}, which are calculated from the same GRMHD data in \fig{fig:grmonty_spec}, but the SED from the non-cooled model (NC3RM; red) is normalised by adjusting the density unit to match the flux of the cooled model (C3D01RM; blue) at 230 GHz (vertical dotted line), which is $\sim$3 Jy. The inclination angle is $60^\circ$--$75^\circ$.}}
%	\label{fig:grmonty_spec_norm}
%\end{figure}
%

The SED in the cooled model (\texttt{C3D01RM}) differs slightly from the non-cooled model (\texttt{NC3RM}): for the cooled model, the overall flux, including the peak value at the sub-mm bump, is slightly lower than for the non-cooled model.
%, and the flux decreases more sharply for increasing frequency after the sub-mm bump.
%The increased flux in the cooled model can be explained by the enhanced magnetic intensity near the BH as a consequence of the decreased disk scale height and the gas pressure, although the accreting plasma is still gas pressure-dominated. 
The near-infrared (NIR) emission originates from the innermost regions ($2\,r_{\rm g} < r < 6\,r_{\rm g}$; see \citealt{Moscibrodzka:09}), where the gas temperature and the magnetic field intensity are high. The relatively weak NIR emission in the cooled model is indicative of the lower gas temperature due to the inclusion of radiative cooling. The peak of the far-UV flux is also slightly higher in the non-cooled model than in the cooled model, mainly due to the higher flux of the seed photons over the NIR-band, and the peak frequency in the non-cooled model is $\sim 3.5$ times larger than in the cooled model. This is because the average increase of energy in the scattering is formulated to $A = 1 + 4\Theta_e + 16\Theta_e^{2}$ \citep{Esin:96}, implying that the higher temperature in the non-cooled model leads to upscattering of photons into the higher energy range. We note that these differences in the SED arise from our adoption of identical density unit values in the GRRT post-processing for both the cooled and the non-cooled models. While the resulting SEDs lie within the observational constraint at 230 GHz (see \fig{fig:grmonty_spec}), if the non-cooled model is normalised in the GRRT calculation by decreasing the density unit to match the flux of the cooled model at 230 GHz, the differences become less significant. The adjusted density unit to achieve this matching of the $230~{\rm GHz}$ fluxes is $~0.84$ times (i.e., smaller than) the cooled model's density unit value, and thus the estimated accretion rate is also smaller in the non-cooled model.

%, which is $\sim$ 3 Jy: the adjusted density unit is 0.84 times lower than the unit for the cooled model, and thus the estimated accretion rate is also reduced. It shows that the effect of radiative cooling on the SED is even less meaningful statistically when we compare the cooled model with the non-cooled model that has the same flux at 230 GHz.}

\begin{figure} 
	\centering
	\includegraphics[width=\columnwidth]{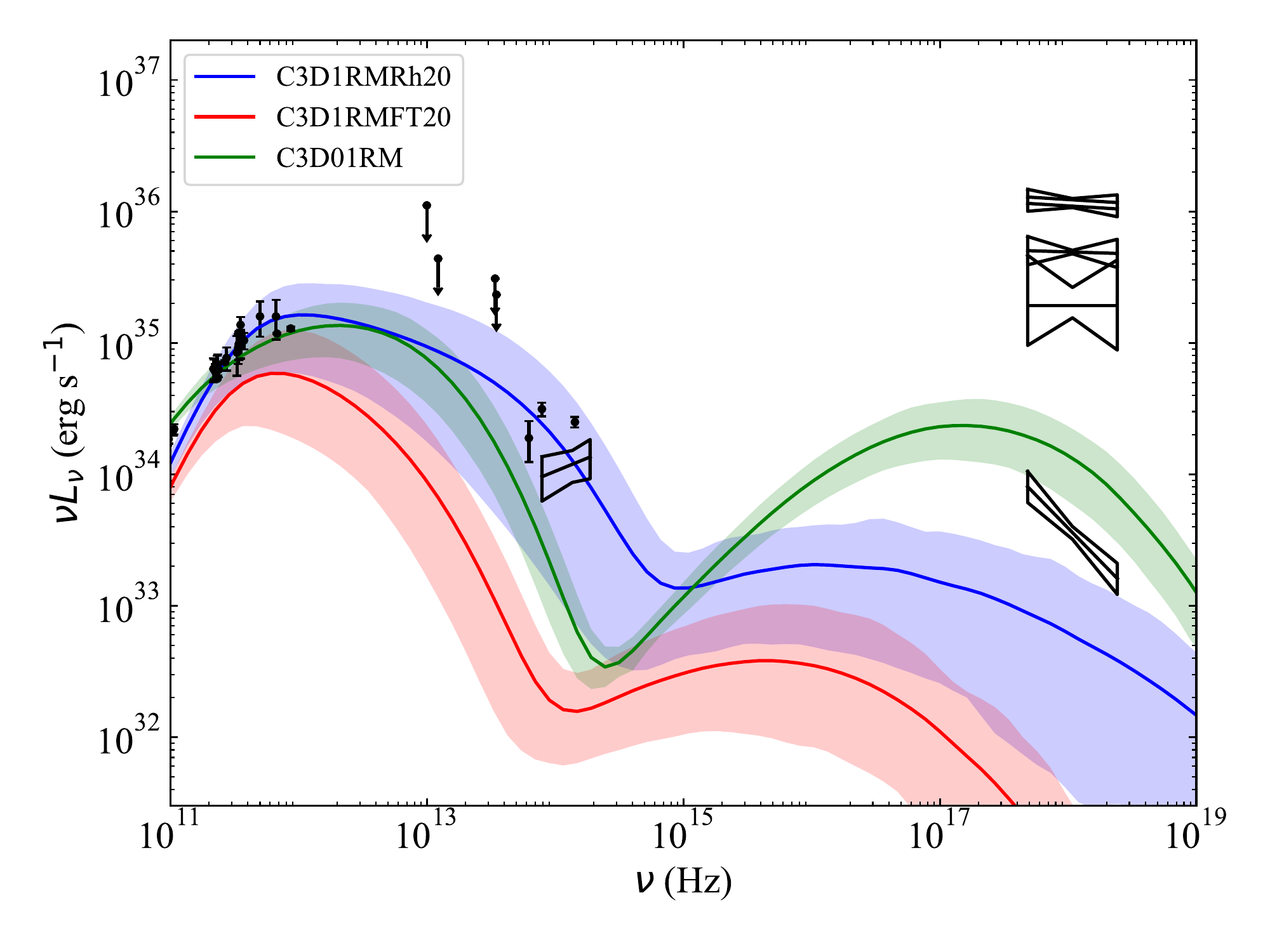}
	\caption{Spectral energy distribution of \sgra{}, calculated using GRMHD results with different electron temperature prescriptions: $T_i/T_e=3$ (\texttt{C3D01RM}, green line), $T_i/T_e=20$ (\texttt{C3D1RMFT20}, red line), and $T_i/T_e$ depending on the plasma magnetisation (\texttt{C3D1RMRh20}, blue line). The inclination angle is $60^{\circ}$--$75^{\circ}$.}
	\label{fig:grmonty_spec_Te}
\end{figure}

Variability studies of Chandra observations showed that $\sim$10\% of the total quiescent X-ray emission likely originates from the inner accretion flow \citep{Wang:13,Neilsen:13}. This indicates that the models can be ruled out in our simulations if they produce X-ray luminosities exceeding $L_{\rm X}\approx2.4\times10^{32}\,\rm erg\,s^{-1}$. We find that for models with the constant ratio of $T_i/T_e=3$, X-ray emission is too strong for both the non-cooled and cooled models at most inclination angles. This implies that the electron temperature should be lower than the value determined by $T_i/T_e=3$. In \fig{fig:grmonty_spec_Te}, we compare the post-processed SEDs from the GRMHD data, which are simulated with a different electron temperature prescription. The model with the increased temperature ratio (\texttt{C3D1RMFT20}; $T_i/T_e=20$) indeed reduces the X-ray emission to below the observed level. However, it is still problematic because its NIR emission is significantly dimmer than the observed values. This may be attributed to the lack of non-thermal electrons in our simulation. Alternatively, a better-fit model can be obtained by adopting an electron temperature prescription that depends on the plasma magnetisation \citep{Moscibrodzka:16, Moscibrodzka:17}, which is expressed as
\begin{equation}
    \frac{T_i}{T_e} = \frac{R_l + \beta^{2}\,R_h}{1+\beta^2}\,,
\end{equation}
where $\beta \equiv P_{\rm gas}/P_{\rm mag}$, and $R_l$ \& $R_h$ are free parameters, which control the dominance of emission depending on the magnetic field strength. The temperature ratio converges into $R_h$ and $R_l$ values at the disk ($\beta\gg 1$) and the jet ($\beta \ll 1)$, respectively. We carry out the simulation with one set of $R_l=1$ and $R_h=20$, and the resulting spectrum is in good agreement with the observed data (shaded blue line in \fig{fig:grmonty_spec_Te}), except for the mismatch of the NIR power-law slope: it reproduces the observed NIR emission while keeping the X-ray emission within the observed maximum limit in the quiescent state. This reinforces the point that SEDs calculated from GRMHD data can be sensitive to the electron temperature prescription, as was investigated recently by \citet{Anantua:20} with a wide parameter space in their ``critical beta'' electron temperature model and equipartition-based constant electron beta/magnetic bias models.  

%We found that in our fiducial setup, the non-cooled model produces too strong X-ray emissions for every inclination angles. For the cooled model, the emission drops sharply at high frequency band after the far-UV bump as a consequence of the lower temperature and the weaker seed photons. It yields that cooled model remains below the critical X-ray value for most inclination angle, while it is marginal in edge-on view. This does not necessarily imply that high inclination angle should be ruled out, but implies that cooling plays a role in reducing the X-ray luminosity and thus it should be considered when one compares the X-ray results between simulation data and observation. Note that other system parameters (e.g., magnetic topology, electron distribution, temperature ratio of electron and ion) may alter the shape of the spectrum. However, we expect that the reduced X-ray luminosity by radiative cooling is a general feature regardless of the changes in the parameters.When the inclination is low ($\theta_{0}=45^\circ$ -- $60^\circ$), X-ray emission from both the cooled model and the non-cooled model are well below the critical value. However, for the high inclination angle ($\theta_0 > 60^\circ$), the non-cooled model is too bright in X-ray band, while the cooled model remains below the critical value.  

\subsubsection{Synthetic Images at 230 GHz}
\label{subsubsec:image}

\begin{figure*} 
	\centering
	\includegraphics[width=\textwidth]{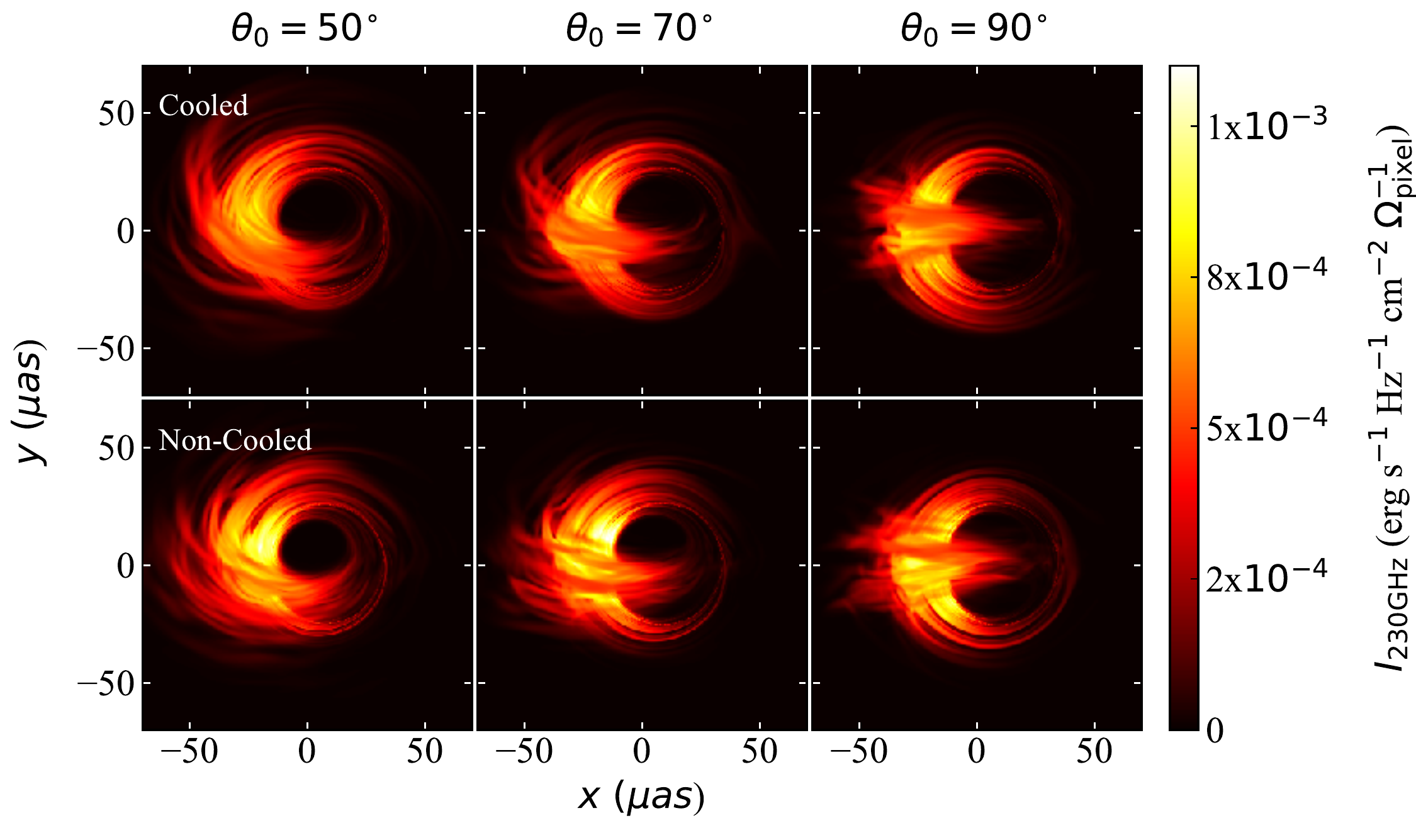}
	\caption{GRRT synthetic images at $230$ GHz for the cooled model (\texttt{C3D01RM}, top row) and the non-cooled model (\texttt{NC3RM}, bottom row). They are taken from a single snapshot at $8000~t_{\rm g}$. The columns represent the result with different inclination angles: $50^\circ$ (left), $70^{\circ}$ (middle), $90^{\circ}$ (edge-on, right). The images are post-processed with the GRRT code, \bhoss{}\citep{Younsi:20}.}
	\label{fig:bhoss_image}
\end{figure*}

To compute the synthetic mm (230 GHz) images, we use the GRRT code \bhoss{} \citep{Younsi:12,Younsi:16,Younsi:20,Younsi:19_polarizedbhoss}, in which the radiation processes include synchrotron emission and absorption. The inclination angle of both the BH spin axis and the disk normal to the line of sight are currently poorly constrained by observations. For example, in kinematic studies of the star S2 near the Galactic Centre, the inclination angle is best-fitted to 134$^\circ$, which is moderate \citet{Gravity:18}.  One may also expect the system to be nearly edge-on given that highly inclined sources often produce linear polarisation in compact radio sources around \sgra{} \citep{Bower:03}. In this work, we set inclination angles of $50^{\circ},\,70^{\circ},\,{\rm and} \,90^{\circ}$ to clearly see how the inclination angle changes the shape of spectra and images.  Note that we assume an aligned jet with the angular momentum of accreting gas.  In fact, misalignment is likely to occur in \sgra{} since infalling gas cannot be quickly torqued into alignment with the BH spin given the geometrically thick disk \citep{Dexter:13,Liska:18a,White:20}.\footnote{Contrary to the thick disk, in thin high-viscosity disks ($H/R < \alpha$, where $\alpha$ is the viscosity parameter) the disk warps are propagated through viscous diffusion and the inner disk aligns with the BH spin axis through {\it Bardeen-Petterson alignment} \citep{Liska:19}.} We will present the effects of tilted accretion disks in a different work \citep{Chatterjee:20}. \fig{fig:bhoss_image} shows the synthetic images at $230$ GHz for a single time snapshot at $8000$ $t_{\rm g}$, which shows detailed turbulent substructure. The ring-like structure is produced by the gravitational lensing effect that magnifies the emissions from accretion flow, and the bright patches in the left-side are the result of Doppler boosted emissions from the approaching side of the disk approaching the observer. These patches are brighter in the non-cooled model than in the cooled model due to the relatively higher gas temperature near the BH. 

\begin{figure*} 
	\centering
	\includegraphics[width=\textwidth]{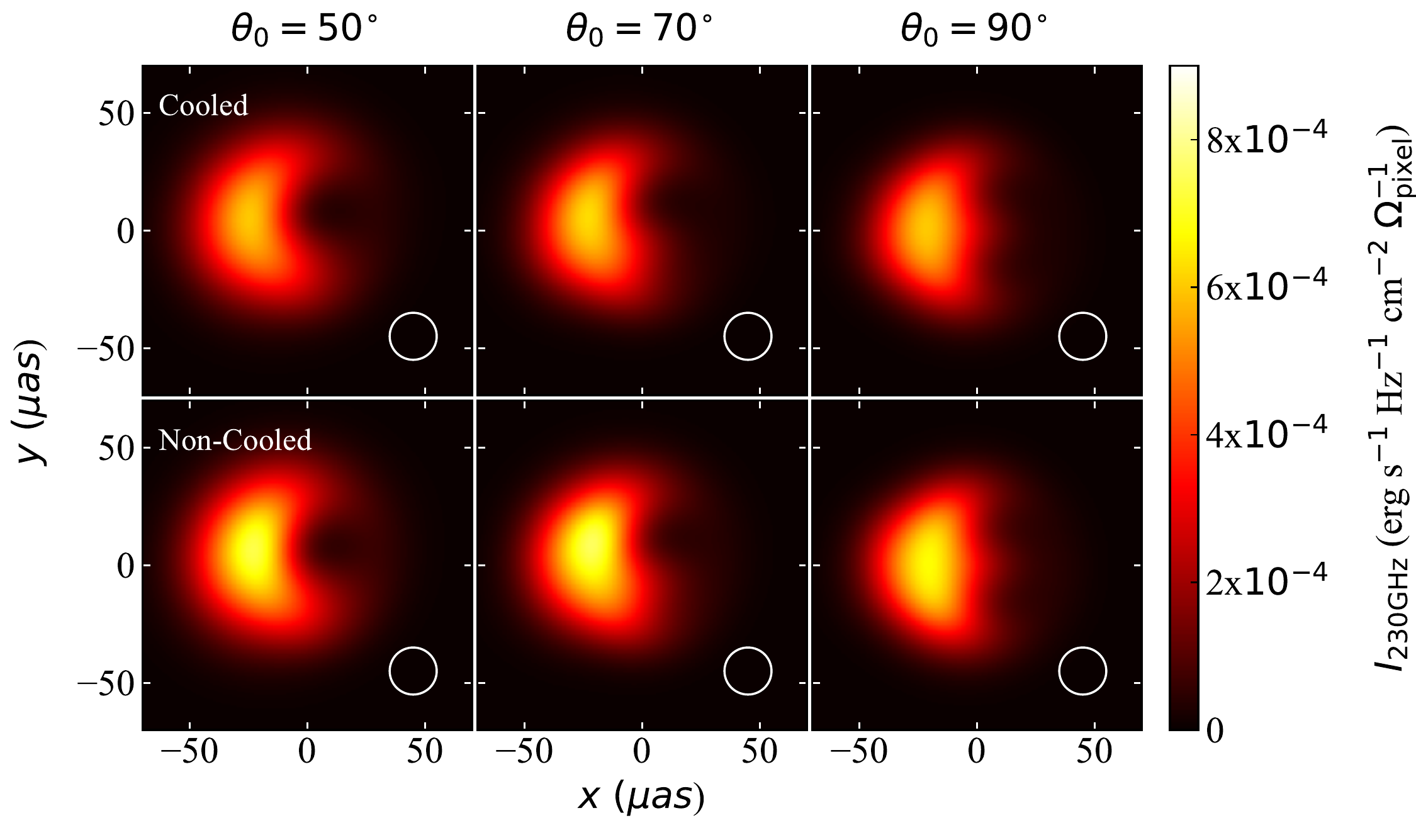}
	\caption{Time-averaged and blurred images at $230$ GHz for the cooled model (\texttt{C3D01RM}, top row) and the non-cooled model (\texttt{NC3RM}, bottom row). The synthetic images are averaged over $8000~t_{\rm g}$--$9500~t_{\rm g}$ and blurred by the convolution with a 2D Gaussian filter, at which the full-with-half-maximum (FWHM) is $20~\mu {\rm as}$ (white circle in bottom right of each panel). The size of the images extends to $\pm 65~\mu {\rm as}$, which corresponds to $\sim 15~r_{\rm g}$. The columns represent the result with different inclination angles: $50^\circ$ (left), $70^{\circ}$ (centre), $90^{\circ}$ (edge-on, right).}
	\label{fig:bhoss_image_blrr}
\end{figure*}

In \fig{fig:bhoss_image_blrr}, we take the blurred images, which are averaged over the time interval between $8000$ -- $9500$ $t_{\rm g}$, as an appropriate proxy for the \eht{} image. The time interval of $1500$ $t_{\rm g}$ corresponds to $\sim8.4$ hrs given the BH mass is $\sim4.1\times10^6\,\MsunYr$ \citep{Gravity:18}. The blurred images were obtained through convolution with a Gaussian filter, at which the full with half maximum (FWHM) is 20 $\mu$as. As seen in the figure, the emission is dominated by the left side of the disk, which has a symmetric crescent shape for all models. Such a crescent shape is the blurred region of the aforementioned hot patches, which are produced by Doppler beaming.  The intensity contrast within the crescent in the non-cooled model is larger compared to the cooled model, as expected from the higher temperature in the regions corresponding to the hot patches.  In general, the BH shadow is clearly visible for lower inclination angles ($\theta_0 < 70^{\circ}$), but becomes less visible for the edge-on images. 
%The emission region is broader in the cooled model than in the non-cooled
%model, which should be considered in the EHT imaging of \sgra{}, although the different parameters
%(e.g., lower $T_i/T_e$) can lead to the similar effect.

\subsection{Variability}
\label{subsec:variability}

\begin{figure} 
	\centering
	\includegraphics[width=\columnwidth]{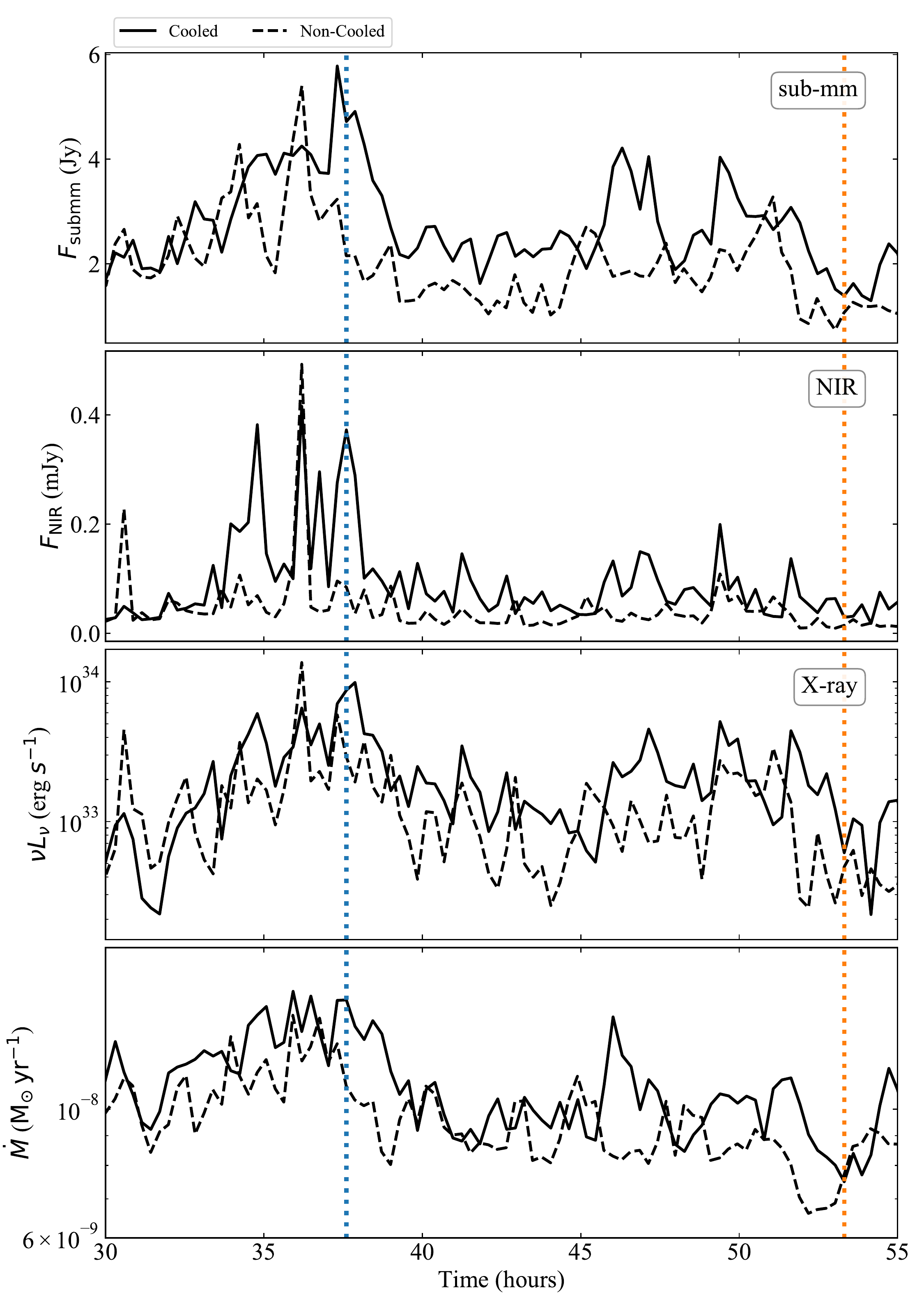}
	\caption{Light curves at three different frequency bands, from top to bottom: sub-mm ($9.4\times10^{11}$ Hz), near-IR ($4.5\times10^{14}$ Hz), and X-ray ($5\times10^{17}$ Hz) for the cooled model (\texttt{C3D1RMRh20}, solid curve) and the non-cooled model (\texttt{NC3RM}, dashed curve). The bottom panel presents the mass accretion rate for the cooled and non-cooled models. The light curves are calculated by using \grmonty{} with the assumption of $D_{\rm Sgr} = 8.2$ kpc, where $D_{\rm Sgr}$ is the distance to \sgra{}. The fluxes are averaged over the theta bin of $60^{\circ}$--$75^\circ$. The bottom panel shows the mass accretion rate at the event horizon. The cadence of the simulation is $50~t_{\rm g}$, which corresponds to $\sim 16$ minutes in \sgra{}. The vertical dotted lines indicate the selected time for a flaring-like event (blue) and a quiescent state (orange).}
	\label{fig:lightcurve3}
\end{figure}

The dynamical environment around \sgra{} drives flares through various mechanisms: sudden electron heating by magnetic re-connection, star-disk interactions, stochastic acceleration, gravitational lensing of ``hot spots'' in the accretion flow, and sudden increases in the mass accretion rate due to the infall of clumps of material \citep[e.g.,][]{Markoff:01,Nayakshin:04,Yuan:04,Trippe:07,Hamaus:09,Gravity:18}.

\begin{figure} 
	\centering
	\includegraphics[width=\columnwidth]{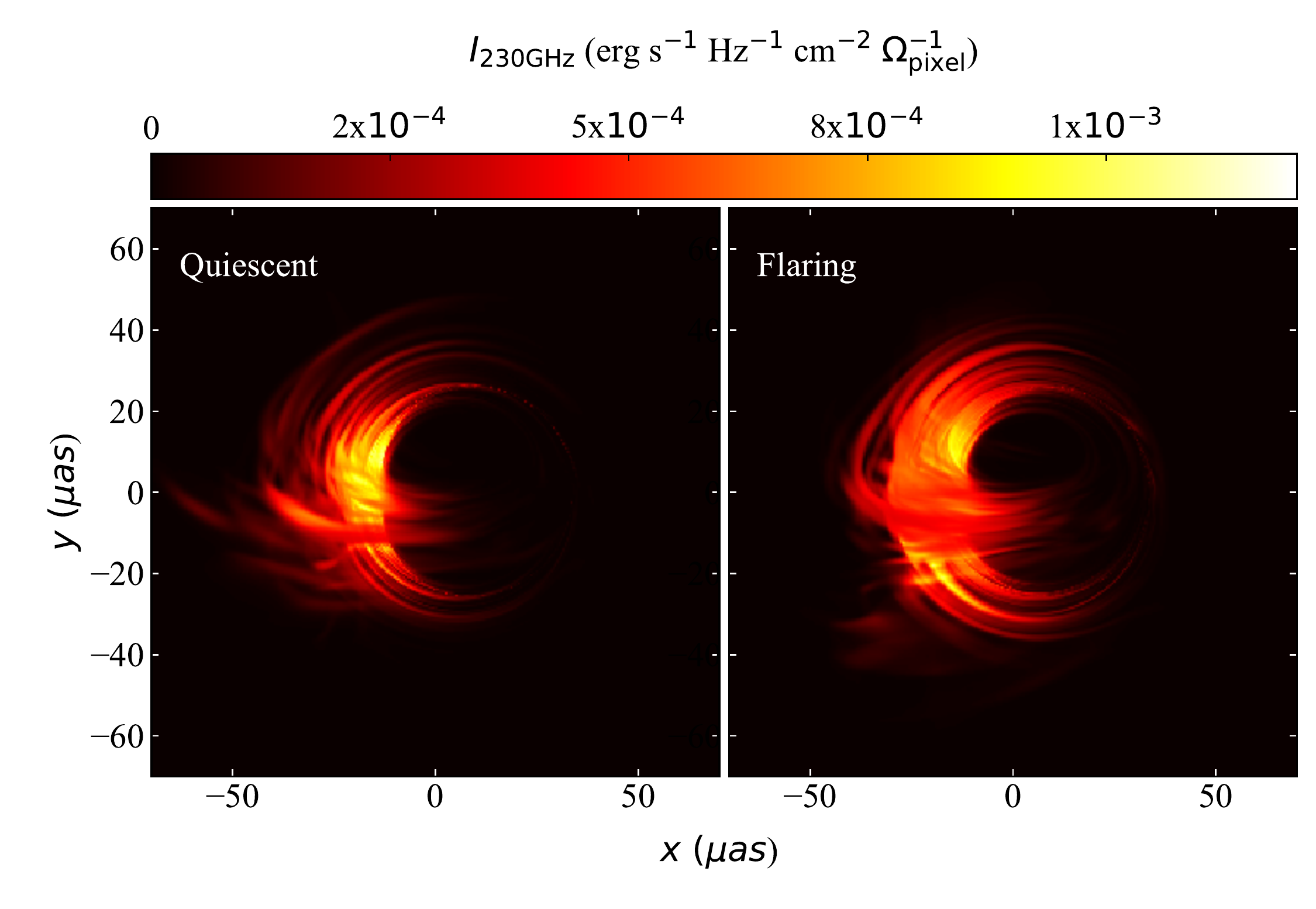}
	\caption{Comparison of GRRT synthetic images at $230$ GHz for the cooled model (\texttt{C3D1RMRh20}) between two different states: NIR-quiescent state at $t=53\,h$ and NIR-flaring state at $t=37\,h$ (see \fig{fig:lightcurve3}). The inclination angle is 70$^\circ$.}
	\label{fig:flare_img}
\end{figure}

\begin{figure} 
	\centering
	\includegraphics[width=\columnwidth]{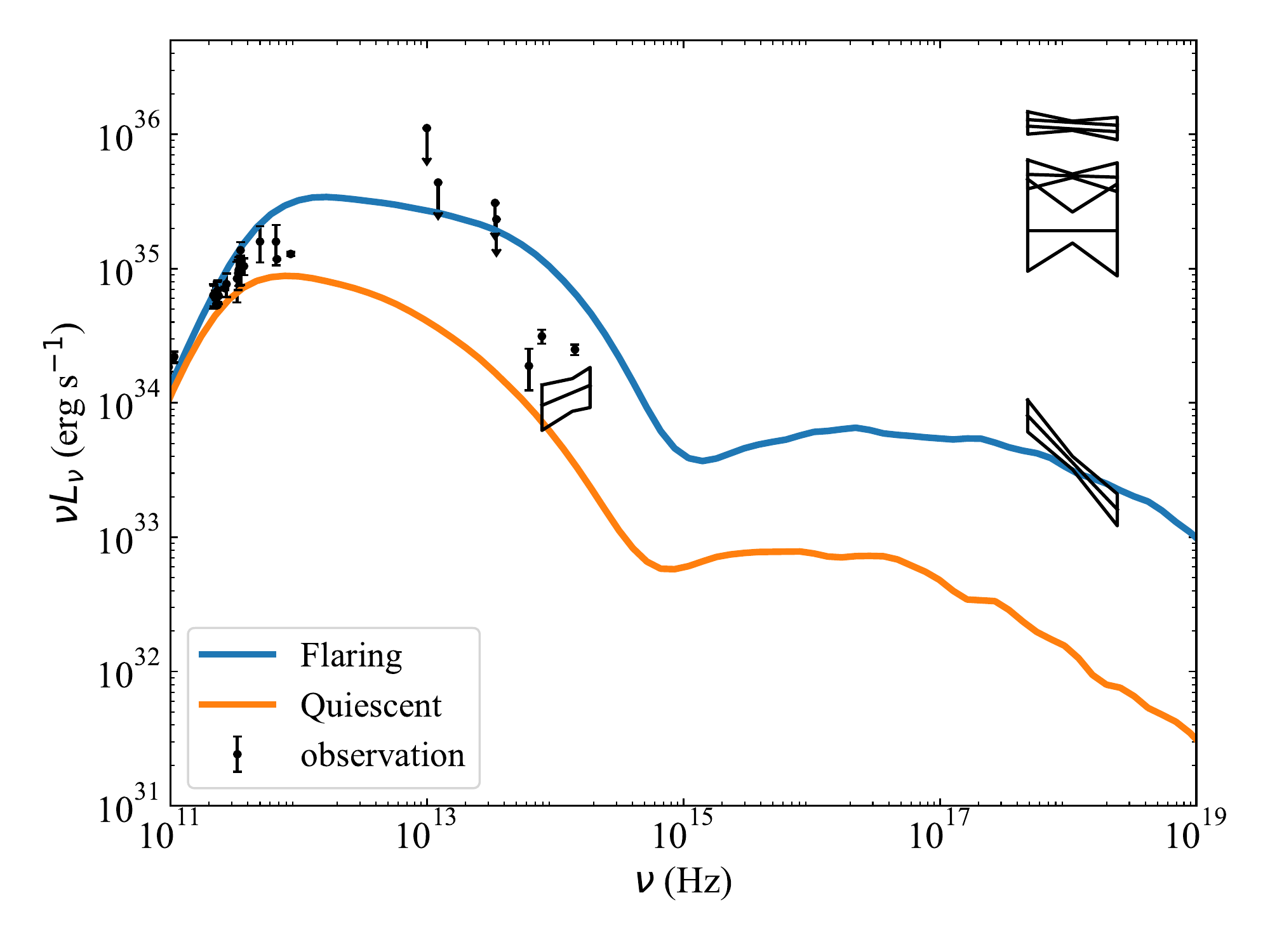}
	\caption{Comparison of SEDs between two different states: NIR-quiescent state at $t=53\,h$ and NIR-flaring state at $t=37\,h$ (see \fig{fig:lightcurve3}). The inclination angle is 70$^\circ$.}
	\label{fig:flare_spec}
\end{figure}

As seen in \fig{fig:lightcurve3}, both the cooled (\texttt{C3D1RMRh20}) and non-cooled (\texttt{NC3RM}) models are highly variable in the different multi-wavelength bands.
%exhibit the flaring events for the multi-wavelength bands. 
The NIR lightcurve behaves similarly to the X-ray lightcurve, and the eruption events in both wavebands are roughly correlated with a pronounced rise in the mass accretion rate. \fig{fig:flare_img} shows the $230$ GHz images for the cooled model \texttt{C3D1RMRh20} during the NIR quiescent state (orange dotted line in \fig{fig:lightcurve3}) and a flaring state (blue dotted line). In this figure, the optical depth of the accretion flow increases during the flaring event, in accordance with the increase in the mass accretion rate. \fig{fig:flare_spec} shows the corresponding SEDs, and clearly illustrates the overall rise in flux across all frequencies. The rise in the X-ray emission is relatively larger than in the NIR, and while the X-ray flux level is close to the quiescent limit, the NIR flux exceeds the quiescent flux by a fact $\sim 4$. Hence, an increase in the accretion rate can probably trigger a NIR flare without a clearly detectable X-ray flare, which can account for why some NIR flaring events do not exhibit simultaneous X-ray flares \citep{Hornstein:07}. This can be seen for both the cooled and non-cooled models, implying that radiative cooling perhaps plays little role in producing flares. However, it is expected that such cooling shortens the duration of the flaring events, which its importance can be significant, in tandem with electron heating (Leichtnam et al., in prep).
%Fig.~\ref{fig:flare_img} \& Fig.~\ref{fig:flare_spec} shows clearly that when the eruption events occur, the emission region becomes broader and overall SED becomes brighter than in the quiescent state, possibly because of the increased mass accretion rate. 

However, the maximum peak of NIR emission is $\sim 0.5$ mJy, which is an order of magnitude lower than the observed flaring flux \citep{Dodds-Eden:11}. Moreover, the X-ray luminosity lies below the value of $2\times 10^{34}\, \rm erg\, s^{-1}$ for the entire time period, which is still identified as an X-ray 'quiescent' state: the observed luminosities of the bright X-ray flares are $>10^{35}\,\rm erg\,s^{-1}$ \citep{Haggard:19}. One possible reason is that our assumption of purely thermal electrons is not sufficient to produce NIR flares, since non-thermal electrons can be produced near the BH via relativistic magnetic re-connection \citep{Werner:16}. The contribution of non-thermal electrons to flaring will be discussed in an upcoming paper (Chatterjee et al., in prep). As an alternative solution, \citet{Dexter:20} suggested that saturation of magnetic flux can trigger the flaring events in a magnetically arrested disk.

\subsection{Comparison with previous works: 2.5D vs. 3D}
\label{subsec:comp2D3D}

\begin{figure} 
	\centering
	\includegraphics[width=\columnwidth]{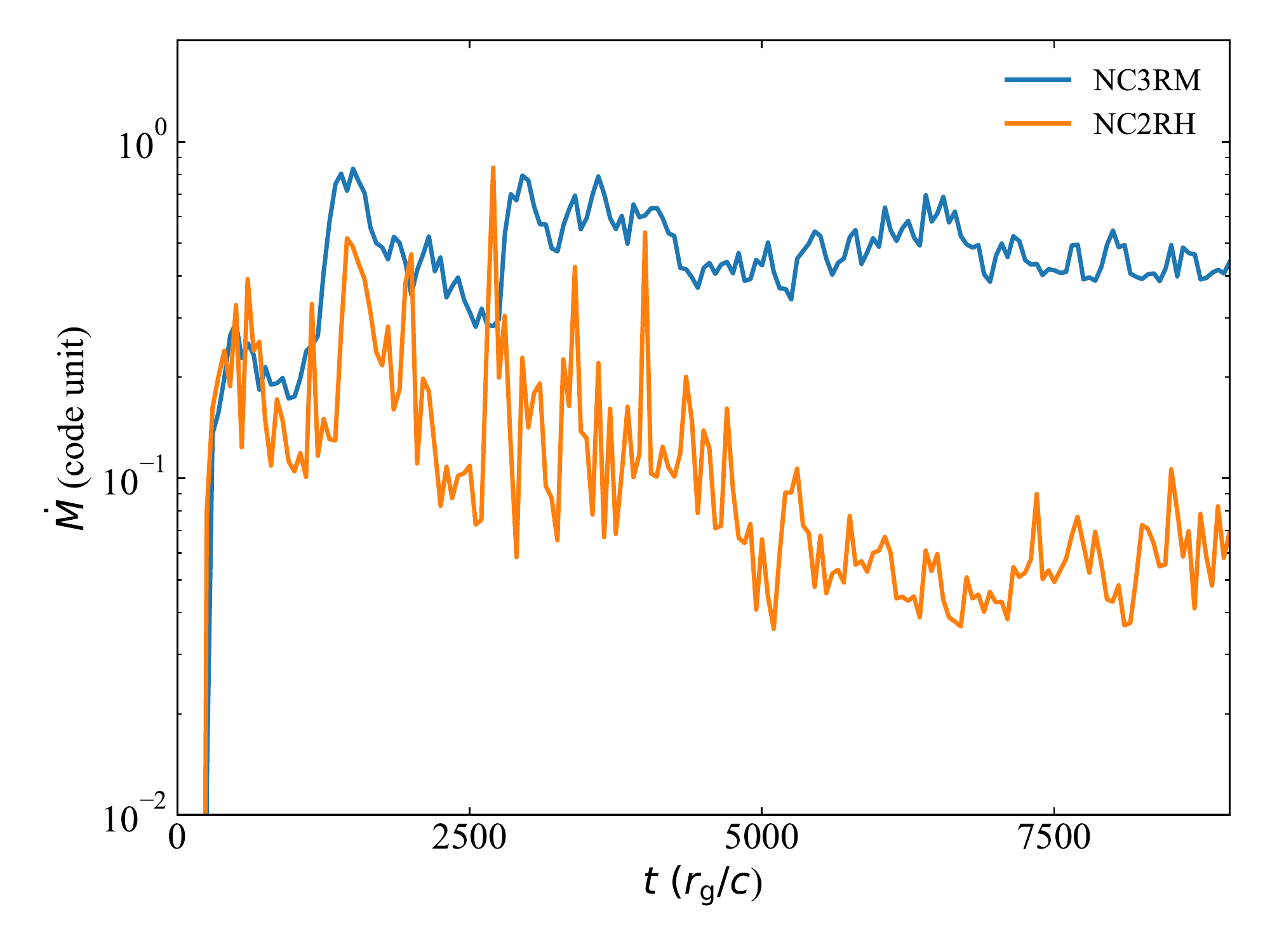}
	\caption{Dimensionless mass accretion rate as a function of time for the non-cooled high resolution 2.5D model (\texttt{NC2RH}, orange) and the non-cooled medium resolution 3D model (\texttt{NC3RM}, blue).}
	\label{fig:accr_2D3D}
\end{figure}

Recent axisymmetric 2.5D GRMHD simulations have explored the effects of radiative cooling on the dynamical evolution of hot accretion flows around \sgra{} \citep[e.g.,][]{Fragile:09,Straub:12,Dibi:12,Drappeau:13}.  However, it is known that MRI-driven turbulence is not sustainable in the axisymmetric 2.5D simulations, which decays over the local orbital time as a consequence of the Cowling's anti-dynamo theorem \citep{Hide:82}.  As seen in \fig{fig:accr_2D3D}, the mass accretion rate in the 2.5D run never reaches a steady-state and significantly decreases after 1000 $t_{\rm g}$ due to the lack of angular momentum transport via the MRI, while the mass accretion rate in the 3D run reaches a quasi-stationary state. Therefore, the 2.5D, axisymmetric approximation does not allow running over a long simulation time, and instead it requires choosing the data at a certain period of time before the MRI decays dramatically, or including an artificial magnetic dynamo term in the induction equation \citep{Sadowski:15}.

Nevertheless, in many respects our 3D results agree with the previous 2.5D results \citep{Dibi:12} in that radiative cooling plays an increasingly significant role with increasing mass accretion rate and its impact becomes important above a mass accretion rate of $\dot{M} > 10^{-8}\,\MsunYr$. The best-fit \sgra{} model with the constant temperature ratio of $T_i/T_e=3$ in our work requires the mass accretion rate of $\sim 10^{-9}\,\MsunYr$, which is similar to the results from the previous 2.5D results \citep{Moscibrodzka:09, Dibi:12}.

%\cmtdy{Working on the comparison of radiative properties (spectra and image) between 2D and 3D data to figure out how the scaling and shape are different}

%\subsection{Implication}
\section{Conclusions}
\label{sec:conclusions}

By means of GRMHD simulations and GRRT post-processing, we study the effects of radiative cooling on the dynamics of accretion flows and their resulting spectra. It is generally assumed that radiative cooling is negligible for RIAF disks, which occur at low accretion rates (i.e., $\dot{M} \lesssim 10^{-7}\, \dot{M}_{\rm Edd}$). However, the importance of radiative cooling increases with the increasing accretion rate.
It is poorly understood what the critical value of the BH mass accretion rate beyond which radiative cooling becomes effective actually is, particularly in 3D. 

For the calculation of radiative cooling, we adopt the approximate solution for the advection-dominated accretion disk, which includes bremsstrahlung, synchrotron, and inverse Compton scattering \citep{Esin:96}. We assume that the temperature ratio between ions and electrons, $T_i/T_e$, is constant or depends on the plasma beta. However, recent studies with particle-in-cell simulations shows that the temperature ratio increases over time due to the weak ion-electron thermal coupling \citep{Zhdankin:20}. While it remains unclear if there are other mechanisms for the effective energy transfer from ions to electrons \citep[e.g., the ion cyclotron instability;][]{Sironi:15}, the subject of changing temperature ratio over time is beyond the scope of the current paper. In this work, full 3D GRMHD simulations with radiative cooling are extended from previous 2.5D simulations studied by \citet{Dibi:12,Drappeau:13}.

In general, radiative cooling enhances the disk mid-plane density as the gas pressure decreases due to energy loss, which is radiated away. The disk with reduced pressure and compressed volume increases the dominance of magnetic fields, which reduce the angular momentum transport outwards via the MRI. As a result, when radiative cooling is on, the disk structure is different from when cooling is neglected: a density peak appears near the central BH, and the distance of the peak from the BH increases with the strength of radiative cooling (i.e., mass accretion rate). This difference is negligible when the accretion rate is small, however, when the accretion rate is larger than $10^{-8}\, \MsunYr$, it becomes apparent (see \fig{fig:dcomp}). Since this rate lies within the range of mass accretion rates for \sgra{}, we argue that cooling losses can affect the dynamical evolution to an appreciable degree. 

The effects of radiative cooling on the spectra are visible even for the low accretion rate of $\dot{M} < 10^{-8} \, \MsunYr$: cooling reduces the peak flux in the sub-mm bumps due to the decreased gas temperature. The decreased seed photon by synchrotron at the sub-mm bumps results in the decrease of the flux in the X-ray bumps, for which inverse Compton is responsible. The synthetic images at $230$ GHz, which is calculated by GRRT post-processing, show similar crescent shapes between the cooled and non-cooled GRMHD data, but slightly dimmer in the cooled data due to the decreased temperature adjacent to the BH. 

Recent studies by \citet{Ressler:20} indicate that the inner regions near the BH could be strongly magnetised, as magnetic fields get advected from stellar winds \citep[see also][]{Ressler:18}, which can lead to the formation of MADs. While it is thought to be inevitable for MADs to produce strong outflows, which are absent in \sgra{} \citep[e.g.,][]{Markoff:07}, we plan to conduct a study of MADs with radiative cooling in future work, so as to investigate how cooling would affect the disk and the outflow in such a situation. Another notable caveat of our simulations is the absence of non-thermal electron acceleration, which is deemed to be responsible for X-ray (and perhaps, near-infrared) flaring in \sgra{} \citep{Neilsen:13, Ball:16, Connors:17}. We will discuss the contribution of the non-thermal electron in a different work (Chatterjee et al.~in prep).

%We did not consider magnetically arrested disks, or MADs, in this work. \citet{Ressler:20} found that the accreted gas from stellar winds orbiting \sgra{} \citep[see also][]{Ressler:18} could, in fact, generate strong magnetic fields via turbulence. However, MADs usually lead to strong outflows, which are apparently absent in Sgr~A$^*$ \citep[see, e.g.,][for a counter view]{Markoff:07}. We plan to perform a study of MADs with radiative cooling in future work. 

\section*{DATA AVAILABILITY}
The data from the GRMHD simulations and GRRT calculations used in this work are publicly available at \href{https://doi.org/10.5281/zenodo.3988208}{https://doi.org/10.5281/zenodo.3988208}.

\section*{Acknowledgements}
\label{sec:acks}

This research was enabled in part by support provided by 
Oak Ridge Leadership Computing Facility, which is a 
DOE office of Science User Facility supported under contract DE-AC05-00OR22725, and Calcul Quebec (\url{http://www.calculquebec.ca}) and Compute Canada (\url{http://www.computecanada.ca}).
%NSF PRAC award no. 1615281 at the Blue Waters sustained-petascale computing project and supported in part under grant no. NSF PHY-1125915. 
DY, KC and SM are supported by the Netherlands Organization for Scientific Research (NWO) VICI grant (no. 639.043.513), ZY is supported by a Leverhulme Trust Early Career Research Fellowship, ML was supported by the NWO Spinoza Prize (PI M.B.M. van der Klis) and AT is supported by Northwestern University and by National Science Foundation grants AST-1815304, AST-1911080.

%\clearpage

%%%%%%%%%%%%%%%%%%%%%%%%%%%%%%%%%%%%%%%%%%%%%%%%%%

%%%%%%%%%%%%%%%%%%%% REFERENCES %%%%%%%%%%%%%%%%%%

\bibliographystyle{mnras}
\bibliography{CoolDisk}

%%%%%%%%%%%%%%%%% APPENDICES %%%%%%%%%%%%%%%%%%%%%

\appendix
\section{Validation of Spectral Calculation with \grmonty}
\label{sec:GRRT}

\begin{figure} 
	\centering
	\includegraphics[width=\columnwidth]{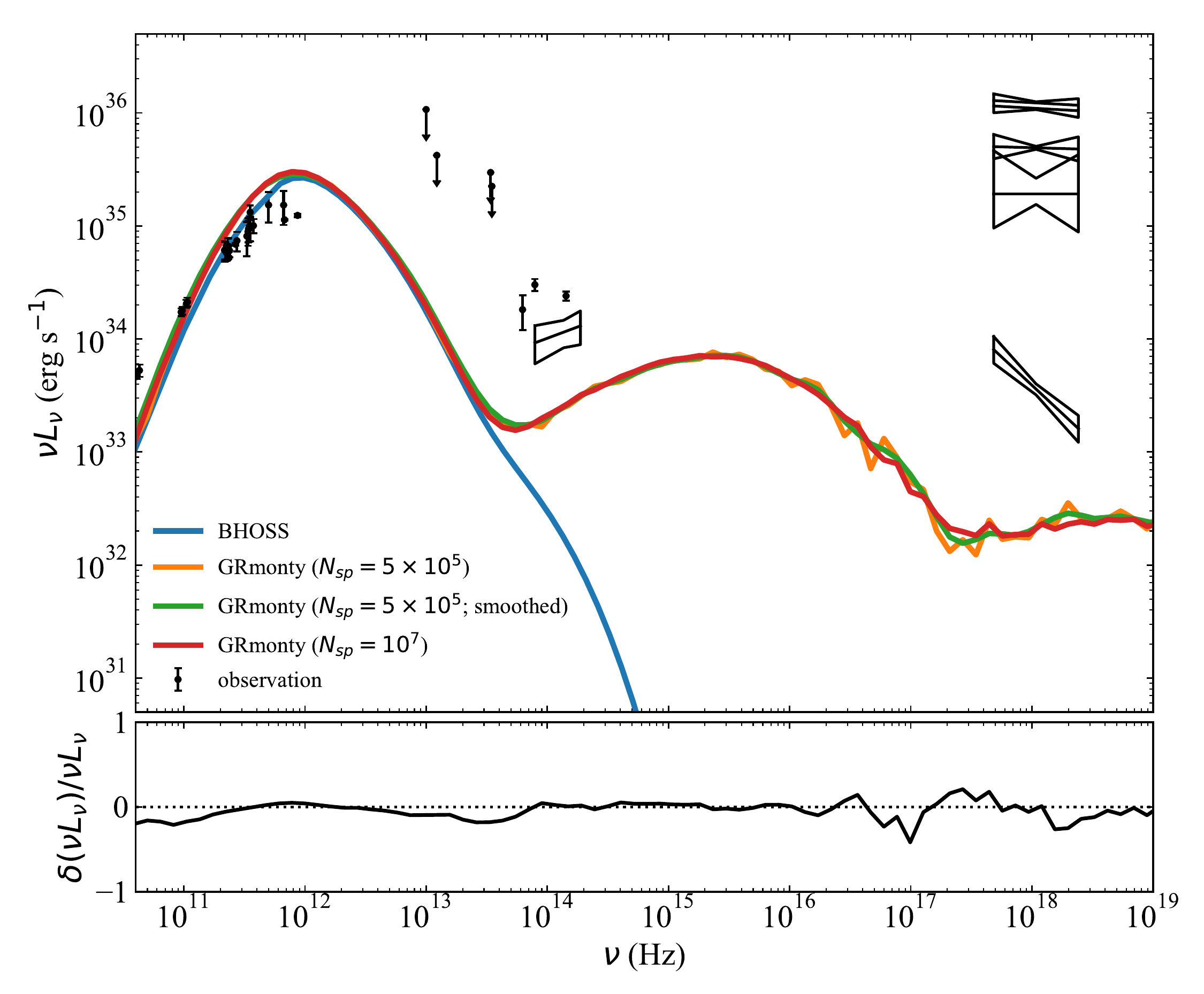}
	\caption{Comparison of spectra calculated in \bhoss (blue curve) and \grmonty. {\it Upper panel}: orange and red curves represent the spectra calculated by \grmonty with $N_{\rm sp}=5 \times 10^{5}$ and $10^{7}$ super-photons, respectively. The green curve represents the smoothed spectrum with $N_{\rm sp}=5\times10^{5}$ super-photons, where a Gaussian filter with $\sigma=1$ is employed. {\it Bottom panel}: fractional difference of spectra between spectrum with a large number of super-photons (red curve) and a smaller number super-photons with the aforementioned smoothing process (green curve), calculated using \eq{eq:devsp}.}
	\label{fig:bhoss_grmonty}
\end{figure}

We make use of \bhoss to reproduce the synthetic images at $230$ GHz, while \grmonty is used to calculate the broadband spectra since the radiative processes in \bhoss include only the synchrotron emission and absorption. To verify consistency between the two codes, we compare the sub-mm bump in the calculated spectra, as the synchrotron emission is dominant in the sub-mm bump. \fig{fig:bhoss_grmonty} shows good agreement for the frequency range of $10^{11}\,{\rm Hz} < \nu < 2\times 10^{13} \,{\rm Hz}$, within which the bump is located.

\grmonty is known to converge to the correct solution as the fractional error $\propto N_{\rm sp}^{-1/2}$ for the optically thin synchrotron sphere, where $N_{\rm sp}$ is the number of the super-photons \citep{Dolence:09}. Evidently, the spectra with different $N_{\rm sp}$ are consistent with each other while the spectra calculated with smaller $N_{\rm sp}$ exhibit more fluctuations at high frequencies than spectra calculated with larger $N_{\rm sp}$. Due to limited computing resources, we choose the number of super-photons as $N_{\rm sp}=5\times 10^5$ for the series of snapshots ($\sim 100$ snapshots for a single run), and to reduce the sampling fluctuations due to small $N_{\rm sp}$, we smooth the spectra using a 1D Gaussian filter with $\sigma=1$. Given that the number of data points is 200, the size of the energy bin is $1.5\times10^{33}\, \rm erg\,s^{-1}$ and the value of $\sigma=1$ corresponds to the FWHM of $3.5\times10^{33}\, \rm erg\,s^{-1}$.
\fig{fig:bhoss_grmonty} shows the difference of the resulting spectra with the different $N_{\rm sp}$. In bottom panel, we calculate the fractional difference, which is expressed as,
\begin{equation}\label{eq:devsp} 
    \frac{\delta(\nu\, L_{\nu})}{\nu\,L_{\nu}} = \frac{ L_{\nu,\rm Nsp7} - L_{\nu,\rm Nsp5, smoothed}}{L_{\nu, \rm Nsp7}}\,,
\end{equation}
where $L_{\nu,\rm Nsp5, smoothed}$ and $L_{\nu,\rm Nsp7}$ are the luminosities, which are calculated with $N_{\rm sp}=5\times10^{5}$ super-photons and a smoothing process, and exclusively with $10^{7}$ super-photons, respectively. As seen in the figure, the fractional difference is small ($\delta(\nu\, L_{\nu})/{\nu\,L_{\nu}} << 1$) for all frequency ranges, and thus we are confident in calculating spectra with $N_{sp}=5\times10^5$. However, this may not be sufficient for the optically thick synchrotron sphere since the correct computation of the photon-weights in the large optical depth regime requires a minimum number of super-photons for convergence \citep{Dolence:09}. The current public release of \grmonty is only available with {\tt Open-MP}, which works with multiple processors in a single node. In future work, especially studies investigating the case of high accretion rates, it is necessary to incorporate acceleration schemes such as {MPI-parallelisation} to be able to use a large number of super-photons.

%%%%%%%%%%%%%%%%%%%%%%%%%%%%%%%%%%%%%%%%%%%%%%%%%%

\bsp	% typesetting comment
\label{lastpage}
\end{document}